\documentclass[%
 reprint,
 amsmath,amssymb,
 aps,
floatfix,
]{revtex4-1}

\usepackage{graphicx}
\usepackage{dcolumn}
\usepackage{bm}
\usepackage{braket}

\begin{document}

\title{Theoretical basis for quantum simulation with a planar ionic crystal in a Penning trap using a triangular rotating wall}

\author{A. Khan}
\affiliation{Department of Physics, Indian Institute of Technology,  Roorkee-247667 Uttarakhand, India}
\author{B. Yoshimura}
\affiliation{Department of Physics, Georgetown University, Washington, DC, 20057 USA}
\author{J. K. Freericks}
\affiliation{Department of Physics, Georgetown University, Washington, DC, 20057 USA}

\date{\today}

\begin{abstract}
One of the challenges with quantum simulation in ion traps is that the effective spin-spin exchange couplings are not uniform across the lattice. This can be particularly important in Penning trap realizations where the presence of an ellipsoidal boundary at the edge of the trap leads to dislocations in the crystal. By adding an additional anharmonic potential to better control interion spacing, and a triangular shaped rotating wall potential to reduce the appearance of dislocations, one can achieve better uniformity of the ionic positions. In this work, we calculate the axial phonon frequencies and the spin-spin interactions driven by a spin-dependent optical dipole force, and discuss what effects the more uniform ion spacing has on the spin simulation properties of Penning trap quantum simulators. Indeed, we find the spin-spin interactions behave more like a power law
for a wide range of parameters.
\end{abstract}

\pacs{37.10.Ty, 03.75.−b, 03.67.Ac, 03.67.Lx}

\maketitle

\section{\label{sec:level1}Introduction}

The idea of a quantum simulator, where a complex many-body quantum system is emulated in a controlled analog quantum computer and the results of the simulation are read off of the computer by measuring different properties as a function of time, originates with work from Richard Feynman~\cite{feynman} in the early 1980s. Cirac and Zoller~\cite{cirac_zoller} showed how ion traps driven by a spin-dependent optical dipole force could realize quantum computers. Porras and Cirac~\cite{porras_cirac} further described how one could perform quantum simulations in a Penning trap. One of the issues with these quantum simulations is that the ions are not spaced uniformly. On the one hand, this leads to nonuniform effective spin-spin couplings between the ions, on the other hand, in one-dimensional linear Paul traps, it leads to the linear to zig-zag transition, which limits the number of ions that can be held in the trap in a one-dimensional linear configuration. It was quickly realized that by adding an anharmonic potential, which pushes together the farther out ions preferentially when compared to the central ions, one can achieve a more uniform arrangement, and the precise potential for perfectly uniform trapping is known for the linear Paul trap~\cite{duan_monroe}. Surprisingly, one can achieve quite uniform crystals by just adding a quartic potential on top of the conventional quadratic trapping potential. This ideology has been extended to the Penning trap by Dubin~\cite{dubin}, where he also included a triangular-shaped rotating wall potential to reduce dislocation formation, which occurs in two-dimensional Penning traps when the boundary potential does not have the same symmetry of the underlying ionic lattice.

In this paper, we extend the analysis of Dubin to determine the behavior of different numbers of trapped ions, different wall potentials and different rotation rates, to determine the stability of these ionic crystals. We further calculate the axial phonon modes and from them the effective spin-spin interactions induced by a state-dependent optical dipole force. We end by discussing the feasibility of the triangular wall for quantum simulation with ionic crystals in the Penning trap.
The organization of this paper is as follows: in Sec.~II, we describe the theoretical background for
the calculations. In Sec.~III, we present the numerical results for the calculations of the spin-spin
couplings of the ions. In Sec.~IV, we present our conclusions.

\section{\label{sec:level2}Theoretical Formulation}

The Penning trap confines ions by using an electrostatic potential that pushes the ions towards the plane with $z=0$, and also pushes the ions outwards, radially. A large static magnetic field curves the radial motion into circles, which result in a trapped ion crystal (after taking into account the Coulomb repulsion
of each ion). An additional rotating wall potential, with specified angular symmetry, is then applied to
control the shape of the crystal and its rotation rate. While many different ions can be employed in
a Penning trap, we will focus on the realization with  two hyperfine levels of ${}^{9}$Be$^{+}$ ions, $\ket{{}^{2}S_{1/2},m_{J}=1/2}$ and $\ket{{}^{2}S_{1/2
},m_{J} = -1/2}$, localized  to a single plane. An extensive description of such a set-up can be found elsewhere~\cite{bollinger_penning_trap}. Cold atoms condensing in such crystals are candidates for building quantum simulators, owing to the ease with which these systems can be prepared for large number of ions, and the precise quantum control of individual ions these systems afford~\cite{britton}.

In actual experiments, the ionic crystal often acquires an increasing number of impurities as a function of time, such as BeH${}^{+}$ that form due to collisions of the beryllium ions with hydrogen molecules that are in the dilute background gas. While a theoretical treatment that includes the effect of these impurities
is possible~\cite{mcaneny_yoshimura_freericks}, we consider only the clean limit here, where there are no impurities. This simplifies the analysis below. In addition, experimental protocols to purify the systems may make this effect less important~\cite{bollinger_photodissociation}.

The theoretical treatment of the equilibrium positions and normal modes of the Penning trap
requires a careful analysis employing standard classical mechanics and then an appropriate quantization scheme. Details for how to do this have already appeared~\cite{russian,taylor,wang_keith_freericks}. Here, we provide a quick summary of that formalism to establish our notation and to show how the approach needs to be modified for the more uniform triangular crystals that can be generated in an extra anharmonic potential with a rotating triangular wall. We begin with the ion Lagrangian in the laboratory reference frame which satisfies
\begin{equation}
\mathcal{L} = \sum_{j=1}^{N} \left[ \frac{1}{2}m\mathbf{\dot{r}_{j}^{2}} - e\phi_{j}(t) + e\mathbf{A}\cdot\mathbf{\dot{r}}_{j} \right]
\label{eq: lagrangian}
\end{equation}
where $N$ is the total number of ions, $e$ is the (positive) unit charge of an electron and $m$ is the mass of a ${}^{9}$Be${}^{+}$ ion. The symbol $\mathbf{r}_{j}=(x_{j},y_{j},z_{j})$ is the position vector for the $j$th ion in Cartesian coordinates, $\phi_{j}(t)$ is the total scalar potential acting on the $j$th ion (including the rotating wall potential), and $\mathbf{A}_{j} = (\mathbf{B}\times \mathbf{r}_{j})/2$ is the vector potential in the symmetric gauge for the uniform axial magnetic field $\mathbf{B} = B_{z}\hat{z}$ with $(B_{z}>0)$. The scalar potential $\phi_j(t)$ includes the potentials that trap the ions  and the mutual Coulomb repulsion between the ions. It can be expressed as follows: 
\begin{eqnarray}
e\phi_{j}(t) = &V_{0}&\left[z_{j}^{2} - \frac{1}{2}\rho_{j}^{2}\right] + \frac{1}{2}  m\omega_{eff}^{2}C_4\rho_{j}^{4}\nonumber \\
&+&  V_{W}\rho_{j}^{3}\cos[3(\theta_{j}+\Omega t)] + \frac{k_{e}e^2}{2}\sum_{k\not=j} \frac{1}{r_{kj}}
\label{eq: scalar_potential}
\end{eqnarray}
where $\omega_{eff} = \sqrt{\omega_{c}\Omega - \Omega^{2} -\frac{eV_0}{m} }$ is the effective trapping frequency in the rotating frame of the crystal with $\omega_c=eB_z/m$ (see below), $V_{0}$ is the amplitude of the static quadratic potential from the Penning trap electrodes, $C_4$ is the strength of the additional fourth-order anharmonic trapping potential (also coming from the Penning trap electrodes), $V_W$ is the amplitude of the triangular rotating wall potential, $\Omega >0$ is the rotating wall angular frequency (which rotates about the $z$-axis), and $k_e$ is the Coulomb force constant. Here, $r_{kj} = |\mathbf{r_{k}} - \mathbf{r_{j}}|$ is the interparticle distance between the $k$th and $j$th ion and is given by
$\sqrt{(x_k-x_j)^2+(y_k-y_j)^2+(z_k-z_j)^2}$, $\rho_j$ is the polar coordinate radius for the $j$th ion and is given by $\rho_j=\sqrt{x_j^2+y_j^2}$, and $\theta_j$ is the polar angular coordinate for the
$j$th ion which is given by $\theta_j=\tan^{-1}(y_j/x_j)$. The rotating wall potential makes the potential $\phi_{j}$ time-dependent in the laboratory frame. The solution for the ion positions is simplified by transforming to the equivalent equilibrium problem in the rotating frame with angular speed $\Omega$ (where the effective trapping potential becomes time-independent). Transforming to the rotating frame, we arrive at the following time-independent rotating-frame potential for the $j$th ion (which is confined to the plane with $z=0)$:
\begin{eqnarray}
e\phi_{j} &=& \frac{1}{2} m\omega_{eff}^{2} [(\rho_{j}^{R})^{2} + C_{4}(\rho_{j}^{R})^{4}] \nonumber \\
&+& V_{W} [ (x_{j}^{R}) ^{3} - 3 x_{j}^{R} (y_{j}^{R})^2 ] + \frac{k_ee^2}{2} \sum_{k\not=j} \frac{1}{r_{jk}^{R}}
\label{eq: potential_rot}
\end{eqnarray}
where $r_{j}^{R} = ( x_{j}^{R},y_{j}^{R},z_{j}^{R} )$ is the transformed set of coordinates for the rotating frame. The $\Omega$-dependence of the effective trapping frequency is due to the way velocities transform for rotating frames, which now includes the effects of potential terms from the centrifugal and Lorentz forces as well.

Comparing Eq.~(\ref{eq: potential_rot}) with a similar expression appearing in Ref.~\onlinecite{wang_keith_freericks}, we see the following differences: (i) there is an additional anharmonic, fourth-order term that causes the ions in the outer regions of the crystal to be pushed in more strongly, and hence counteracts some of the inhomogeneities that occur due to increasing interion distances as we move outwards; (ii) the angular shape of the rotating wall term has now been adjusted to an $l=3$ angular harmonic as the crystal condenses into a triangular lattice and there is less frustration at the edges if the rotating wall has a symmetry that matches that of the underlying crystal facets~\cite{dubin}. We note that the additional/modified terms retain their form under a transformation from the laboratory frame to the rotating frame. We also remark that the unconventional choice of $\frac{1}{2}  m\omega_{eff}^{2}C_4\rho_{j}^{4}$ (with the coefficient dependent on $\Omega$) for the anharmonic term was made in anticipation of the simpler and rather standard form we get in Eq.~(\ref{eq: potential_rot}). 

The ions in a Penning trap crystal do not always crystallize in a two-dimensional plane. The following approximate criterion is usually required to be satisfied for a planar only configuration:
\begin{equation}
\frac{2eV_0}{(eB_z\Omega - m\Omega^2 -eV_0)}\gg 1
\label{eq: stability}
\end{equation}
which basically ensures that the restoring force in the axial direction is several orders of magnitude bigger than in the radial direction, and the crystal lies in the $z=0$ plane only.

\begin{figure}[ht]
\centering
\hspace{-10mm}
\includegraphics[scale=0.50]{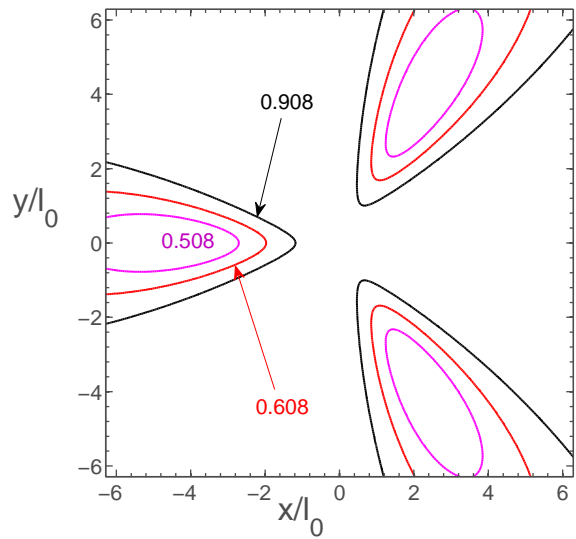}
\caption{(Color online.) Contour plot of the magnitude of the radial restoring force $F_{r}=0$, for parameterized values of the strength of the rotating wall potential (given by $3V_W/\omega_{eff}^2=0.508$, 0.608, or 0.908). The values of this strength have been indicated by appropriately colored labels near each curve. The separatrix radius is roughly the shortest distance to the contour, from the center of the graph and we see that the separatrix radius decreases as we increase the strength of the rotating wall term. \label{fig: separatrix}}
\end{figure}

The triangular rotating wall also introduces an anisotropy in the radial potential, with deconfinement along certain directions which, if strong enough, can lead to particle loss. For the more familiar case of $l=2$, this is reflected in the increasing eccentricity of the elliptical equilibrium structures that suggests deconfinement along the "weak" axis. This is expressed in terms of an approximate criteria in Ref.~\onlinecite{wang_keith_freericks} for the simplest rotating wall. For $l=3$, this becomes difficult to express in terms of a single criterion and we resort to a numerical calculation to find the triangular "separatrix", the contour lines of the radial part of the gradient of the potential function in Eq.~(\ref{eq: potential_rot}), with the Coulomb repulsion terms ignored. These are shown for varying ratios of strengths of the rotating wall term to the effective radial confinement strength, $V_{W}/\omega_{eff}^2$ in Fig.~\ref{fig: separatrix}. We notice that unlike the quadrupole rotating wall potential, deconfinement happens for the triangular wall at large enough distances for all wall amplitude strengths. We see that as the strength of the triangular rotating wall increases, the separatrix moves closer to the center of symmetry of the crystal, with apparent deconfinement centered along the $\theta=0, 2\pi/3 $ and $ 4\pi/3$ axes. This has a noticeable effect on the "shape" of the Penning-trap crystals, which reduces the dislocations in the crystal and helps maintain the uniformity.
Of course, a full description of stability of these planar structures requires inclusion of the Coulomb terms. The accurate quantitative description of stability requires solving for the equilibrium positions presupposing a planar arrangement of ions, and then showing that the eigenvalues of the normal modes of oscillation about these equilibrium positions are all positive; that is, the phonon normal mode frequencies are all real.

The full, transformed Lagrangian for the rotating frame is
\begin{equation}
\mathcal{L}^{R} = \sum_{j=1}^{N} \left[ \frac{1}{2} m\mathbf{|\dot{r}_{j}^{R}|}^2 -\frac{eB^{eff}(\Omega)}{2}(\dot{x}_{j}^{R}y_{j}^{R} - \dot{y}_{j}^{R}x_{j}^{R}) - e\phi_{j}^{R} \right]
\label{eq: lagrangian_rot}
\end{equation}
where $B^{eff}(\Omega) = B_z - 2\Omega m/e $ is the $\Omega$-dependent effective magnetic field in the rotating frame. The modification of the magnetic field is due to velocity dependent terms in the laboratory frame Lagrangian. This affects the oscillating normal modes of the planar motion when the ions are far from their equilibrium positions. However, as we will see soon, this does not have an effect on the axial modes. This observation greatly simplifies the normal-mode analysis for the axial modes.


To find the stable spatial configuration of the ions, we minimize the effective potential energy in the rotating frame of reference. This is a challenging optimization problem to solve for in two (and higher dimensions), especially since different configurations, separated by large potential barriers, can have local minima in the potential energy function with small energy differences to the global minimum. We follow the previous treatments of this problem, where the experimental indication of the fact that the ions condense in a triangular lattice in a single plane is used to construct the optimized solution that lies close to a perfect triangular lattice.

We construct an initial, trial solution based on the ``closed-shell" approximation as in Ref.~\onlinecite{wang_keith_freericks}, but with the important difference that the overall ``shape" is triangular and not hexagonal, as it was in the previous solutions. This is to reflect the fact that the overall shape of the crystal is dictated by the equipotential lines of the rotating wall term, which in this case is triangular. 

We then proceed to calculate the  collective normal mode excitations of the crystal. The ion Lagrangian (in the rotating frame, all $R$ superscripts are dropped for clarity)  is expanded via a Taylor series about the previously calculated equilibrium positions of the ions up to quadratic order. The ion coordinates are, for the purpose of the expansion, written as $\mathbf{r}_{j}(t) = \mathbf{R_{j}^{0}} + \delta \mathbf{R_{j}}(t)$, while for the ion velocities we write $\mathbf{\dot{r_{j}}}(t) = \delta\mathbf{\dot{R_{j}}}(t)$.
Because we are expanding about equilibrium, we can drop the linear terms in the coordinates, and we find
\begin{equation}
\mathcal{L} = \mathcal{L}_0 + \frac{1}{2}\sum_{j=1}^{N} \left.\left[  \delta \mathbf{R}_{j}\cdot\frac{\partial}{\partial \mathbf{R}_{j} } + \delta \mathbf{\dot{R}_{j}}\cdot\frac{\partial}{\partial \mathbf{\dot{R_{j}}}} \right]^2 \mathcal{L}\,\, \right|_{0}
\label{eq: lagrangian_quadratic}
\end{equation}
where the $\mathcal{L}_{0}$ is due to the equilibrium state and the quadratic terms are due to fluctuations away from equilibrium, which we henceforth call $\mathcal{L}_{ph}$ for the phonon Lagrangian. The Lorentz force due to the external magnetic field lies in the $xy$ plane, and the potential energy $\phi_{j}$ is clearly seen to be separable in cylindrical coordinates. This means that the axial phonon Lagrangian can be decoupled from the planar phonon Lagrangian, and there is no harmonic coupling between the planar and axial degrees of freedom. Therefore, we can study the axial and planar modes independently $(\mathcal{L}_{ph}=\mathcal{L}^{axial}_{ph}+\mathcal{L}^{planar}_{ph}$ and we can solve just the equations of motion for the axial or the planar modes independent of the other).

We examine only the axial modes in this work. This is due to the fact that the planar modes have a complex structure owing to a coupling of the ion motion in the $x$ and $y$ directions, the appearance of velocity-dependent forces, as well as complexities introduced by rotation of the ion crystals as observed in the laboratory frame. The fact that there is no harmonic coupling between axial and planar directions of the crystal allows us to exclude the planar modes from our discussion henceforth. Restricting to the axial modes only, is further supported by the fact that the simplest form of quantum simulation works on driving the axial modes with a state-dependent optical dipole force.

The  axial Lagrangian is then given by
\begin{equation}
\mathcal{L}_{ph}^{axial} = \frac{1}{2} \sum_{k=1}^{N}m\left ( \delta \dot{R}_{k}^{z} \right )^2 - \frac{1}{2}\sum_{j,k=1}^{N} K^{zz}_{jk} \delta R_{j}^{z} \delta R_{k}^{z}
\label{eq: lagrangian_axial}
\end{equation}
where the spring constants satisfy
\begin{equation}
K_{jk}^{zz} = - \left.\frac{\partial^{2} \mathcal{L}} {\partial R_{j}^{z} \partial R_{k}^{z} }\right|_{0}
\end{equation}
The absence of cross terms in the velocity part of the Lagrangian can be easily seen from the sum-of-squares structure of the kinetic energy along the $z$-direction and the fact that there are no velocity-dependent forces in the $z$-direction.

An explicit calculation for the matrix elements of the $\mathbf{K^{zz}}$ gives the following:
\begin{equation}
K_{jk}^{zz} = \begin{cases} 2eV_{0} - k_e e^2 \sum_{k',k'\not=j}^{N} \frac{1}{(R_{jk'}^{0})^{3}} & j=k \\
k_e e^2\frac{1}{(R_{jk}^{0})^{3}} & j\not=k \end{cases}
\end{equation}
where $R_{jk}^{0} = |\mathbf{R}_{j}^{0} - \mathbf{R}_{k}^{0}|$ is the distance between ions located at their respective equilibrium positions in the rotating frame. We see that the axial stiffness matrix is Hermitian and symmetric, and is independent of the anharmonic or wall potentials.


To solve for the axial ion normal modes, we apply the Euler-Lagrange equations to the axial phonon Lagrangian in Eq.~(\ref{eq: lagrangian_axial}):
\begin{equation}
m\delta \ddot{R}_{j}^{z} + \sum_{k=1}^{N} K_{jk}^{zz}\delta R_{k}^{z} = 0, \,\,\,\,\,\, j=1,2,....N.
\end{equation}
which, on substitution of the eigenvector solution {\it ansatz} $\delta R_{j}^{\nu}(t) = b_{j}^{z\nu}\cos[\omega_{z\nu}(t-t_0)]$, gives
\begin{equation}
\sum_{k=1}^{N} \,[m\omega_{z\nu}^{2}\delta_{jk} - K_{jk}^{zz}]~b_{k}^{z\nu} = 0, \,\,\,\,\,\, j,\nu=1,2,\ldots N
\end{equation}
where $\omega_{z\nu}$ is the normal mode eigenfrequency and $b_{k}^{z\nu}$ is the $\nu$th axial normal-mode eigenvector. The eigenvalue problem is quadratic, but we can easily map it onto a linear eigenvalue problem by setting the eigenvalue according to  $\lambda^{z\nu} = m\omega_{z\nu}^{2}$. We can then solve for the eigenvalues and eigenvectors numerically in MATLAB. The eigenvectors $b_{j}^{z\nu}$ are real, N-tuples whose norm has been set to unity by convention. The eigenvalues $\lambda^{z\nu}$ are positive for stable normal modes and negative for unstable normal modes.

The quantization of the normal modes is completely standard: we first identify the positions $Q_{\nu}$ and momenta $P_{\nu}$ associated with each phonon mode as canonically conjugate, and promote the relation given by the Poisson bracket $\{Q_{\nu},P_{\nu'}\}=\delta_{\nu\nu'}$ to the commutation relation for the operators $\hat{Q}_{\nu}$ and $\hat{P}_{\nu'}$, $[ \hat{Q}_{\nu},\hat{P}_{\nu'} ] = i\hbar\delta_{\nu\nu'}$. To calculate the canonically conjugate variables for the phonon modes, we make the transformation $\delta R_{j}^{z}(t) = \sum_{\nu} \xi_{\nu}(t) b_{j}^{z\nu}$ where $\xi_{\nu}$ are the normal coordinates for each phonon mode $\nu$. We see that the Lagrangian assumes the following diagonal form:
\begin{equation}
\mathcal{L}_{ph}^{axial} = \frac{1}{2}\sum_{\nu=1}^{N} m (\dot{\xi}_{\nu}^{2} - \omega_{z\nu}^{2}\xi_{\nu}^{2}).
\end{equation}
Hence, we calculate the conjugate momenta as follows:
\begin{equation}
P_{\nu}^{axial} = \frac{\partial \mathcal{L}_{ph}^{axial}}{\partial \dot{\xi_{\nu}}} = m\dot{\xi}_{\nu}.
\end{equation}
The Hamiltonian is then expressed as
\begin{equation}
\mathcal{H}_{ph}^{axial} = \sum_{\nu=1}^{N}\left( \frac{(P_{\nu}^{axial})^{2}}{2m} + \frac{1}{2} m\omega_{z\nu}^{2}\xi_{\nu}^{2} \right) .
\end{equation}
To quantize the normal modes, we identify that the Hamiltonian $H_{ph}^{axial}$ is a sum of simple harmonic modes with frequencies $\omega_{z\nu}$. We now introduce creation and annihilation operators as follows:

\begin{equation}
\hat{a}_{z\nu} = \sqrt{\frac{m\omega_{z\nu}}{2\hbar}} \left( \xi_{\nu} + \frac{i}{m\omega_{z\nu}} P_{\nu}^{axial} \right)
\end{equation}
and
\begin{equation}
\hat{a}^{\dagger}_{z\nu}= \sqrt{ \frac{m\omega_{z\nu}}{2\hbar}} \left( \xi_{\nu} - \frac{i}{m\omega_{z\nu}} P_{\nu}^{axial} \right).
\end{equation}

Hence, the quantized Hamiltonian operator can be expressed as
\begin{equation}
\hat{H}^{axial}_{ph} = \sum_{\nu=1}^{N}  \hbar\omega_{z\nu} \left( \hat{n}_{z\nu} + \frac{1}{2} \right)
\end{equation}
where $\hat{n}_{z\nu} = \hat{a}^{\dagger}_{z\nu}\hat{a}_{z\nu}$ is the number operator. The operator for displacement along the $z$-direction can be expressed in terms of the creation and annihilation operators as follows:
\begin{equation}
\delta \hat{R}_{j} = \sum_{\nu =1}^N b_{j}^{z\nu} \sqrt{\frac{\hbar}{2m\omega_{z\nu}}} \left [ \hat{a}^{\dag}_{z\nu} + \hat{a}_{z\nu}\right ] .
\end{equation}
We also note that the form of the Hamiltonian derived for the axial modes here is invariant when we transform coordinates from the lab to the rotating frame, since the ion oscillations are only along the rotation axis ($z$-direction) and hence are not influenced by rotation of the coordinate axes.

We now need to calculate the effective spin-spin coupling between the ions, generated by the spin-dependent optical dipole force. This analysis has been done in detail elsewhere~\cite{porras_cirac,monroe,wang_freericks}, which we utilize here. The effective spin Hamiltonian is dictated by a time-dependent Ising spin Hamiltonian, 
\begin{equation}
\mathcal{H}(t) = \sum_{j,j'=1}^{N} J_{jj'}(t)\sigma_{j}^{z}\sigma_{j'}^{z},
\label{eq: top}
\end{equation}
where the Ising spin-spin coupling between sites $j$ and $j'$ is given by
\begin{eqnarray}
J_{jj'}(t) &=& \frac{F_O^2}{4m} \sum_{\nu=1}^{N}  \frac{b_{j}^{z\nu}b_{j'}^{z\nu}}{\mu^2-\omega_{z\nu}^{2}} \Biggr [  1+\frac{}{ {}_{}^{} } \cos(2\mu t) \nonumber \\  
&-& \left. \frac{2\mu}{\omega_{z\nu}} \sin \omega_{z\nu}t \sin \mu t \right].
\label{eq: jij}
\end{eqnarray}
Here $F_{O}$ is the magnitude of the optical dipole force, and $\mu$ is the beat-note frequency corresponding to the frequency difference of the two off-resonant laser beams being applied to the trapped ion crystal. We see that this expression relates the strength of the Ising-like coupling between ions to the phonon mode properties $\omega_{z\nu}$, $b_{j}^{z\nu}$ and $b_{j'}^{z\nu}$, which are calculated from the classical, normal-mode analysis described above. The 
time-averaged spin-spin couplings are given by the first term in Eq.~(\ref{eq: jij}). We can think of the effective spin-spin Hamiltonian as the expression in the parenthesis of Eq.~(\ref{eq: top}) with the time-dependent spin-spin interactions replaced by the time-averaged ones.

\section{\label{sec:level3}Numerical Results and Analysis}


We consider ${}^{9}$Be${}^{+}$ ions localized in a plane by the Penning trap potential defined in Eq.~(\ref{eq: scalar_potential}). We characterize the strength of the end-cap potentials $V_0$ that affect the axial trapping by a characteristic angular frequency $\omega_{z}$, such that $eV_0 = \frac{1}{2}m\omega_{z}^{2}$. This is fixed for the purpose of all our numerical calculations at the value $\omega_{z}= 2\pi \times 795$~kHz, a typical value used in experiments. We normalize subsequent frequencies in terms of $\omega_{z}$. The experiments at NIST typically run at rotational frequencies $\Omega = 0.0579 \omega_{z}$, and we have concentrated on regions close to this value in our calculations for experimental relevance. The cyclotron frequency $\omega_{c}$ associated with the magnetic field is defined as $\omega_{c} = eB_{z}/m$. Fixing $B_z = 4.5$~T, we get $\omega_{c} = 9.645\omega_z$. The beryllium atom has an atomic mass $m = 9.012182$~a.u.~and a positive unit charge $e = 1.60217646 \times 10^{-19}$~C.

For the strength of the anharmonic term, we use the value in Ref.~\onlinecite{dubin}, where $\tilde{C}_{4} = 1$ in the following form of the potential $\varepsilon$:
\begin{eqnarray}
\varepsilon  &= &\sum_{i=1}^N \left[\, \frac{1}{2}m\omega_{eff}^2 \left(\rho_{i}^{2} + \frac{3}{8}\tilde{C}_{4}\frac{\rho_{i}^{4}}{r_{p}^{2}}\right)  + \frac{k_ee^2}{2}\sum_{j,i\not=j} \frac{1}{r_{ij}} \right. \nonumber \\
&+& \left. \frac{V_{Wall}^{}}{r_p^3} \rho_{i}^{3}\cos(3\theta_{i}) \right]
\end{eqnarray}
where $r_p$ is the plasma radius parameter whose value is taken to be $r_p =  0.01049$~cm. We now define a typical length and energy scale,
\begin{equation}
 l_0 = \left (\frac{k_{e}e^2}{m\omega_{z}^{2}}\right )^{\frac{1}{3}} ~\mbox{  and  } ~E =m\omega_{z}^{2}l^2_0.
\end{equation}
Henceforth, we express all lengths in units of $l_0$ and all energies in units of $E$.  The potential then becomes
\begin{eqnarray}
\tilde{\varepsilon} =\frac{\varepsilon}{E}&=& \sum_{i=1}^N\left[ \frac{1}{2}\omega_{eff}^{2}(\rho_{i}^{2} + C_4 \rho_{i}^{4}) + \frac{1}{2}\sum_{i\not=j}\frac{1}{r_{ij}} \right .\nonumber\\
&+& V_W (x_{i}^{3}-3x_{i}y_{i}^{2}) \Biggr ]
\label{eq: potential_dimensionless}
\end{eqnarray}
where the dimensionless parameters $C_4$ and $V_W$ are given by the relations $C_{4}=3l^2_0\tilde{C_4}/(8r_p^2)$ and $V_{W}=V_{W} l_0/(m\omega_{z}^{2}r_{p}^{3})$. All lengths ($l_0$), energies ($E$) and frequencies ($\omega_z$) appearing in Eq.~(\ref{eq: potential_dimensionless}) are expressed in dimensionless units. For the choice of $\tilde{C_4}=1$, we have $C_4=0.002472$. For the strength of the wall potential, we use two different values, $V_{W}^{l} = 0.0025$ and $V_{W}^{h}=0.0040$. These have been chosen to show clearly the effect of variation in the wall strength on key ion-crystal characteristics, while also ensuring that all crystal structures have a stable equilibrium.

We report the rotational frequency $\Omega$ in terms of $\omega_{eff} = \sqrt{\omega_{c}\Omega - \Omega^2 - 1/2}$, normalized with respect to $\omega_{z}$. We stick to regions close to the experimental value of $\Omega =  0.0579 \omega_{z}$, which translates as $\omega_{eff} = 0.2339\omega_{z}$. We use values for $\omega_{eff}$ in the range $0.21\omega_z - 0.25\omega_{z}$ in our analysis. 

It is also useful to consider the stability of the crystal under the trap potentials we have used. Because the rotating wall potential varies as the third power of the coordinates, there is no deconfinement frequency as in the case of the quadrupole wall, $l=2$ trap. To probe the stability under deconfining forces, we look at the radial component of the force on the $i$th ion due to the trap potential, excluding the Coulomb potential. which is given by
\begin{equation}
F_{r} = -\omega_{eff}^{2}\left[ \rho_i + \frac{2C_{4}}{\omega_{eff}^{2}}\rho_{i}^{3} + \frac{3V_{W}}{\omega_{eff}^{2}\rho_{i}}(x_{i}^3-3x_iy_{i}^{2}) \right].
\end{equation}

If we plot the locus of points where this function becomes zero for various values of $V_{W}$, we see that we get three regions, each centered along the $\theta_{i} = -\pi/3, \pi/3 \mbox{ and } \pi$ axes. For increasing strength of the rotating wall potential, they move closer to the origin and the radius of the separatrix (the smallest distance to these unstable zones along any axis) is seen to decrease. Hence, the deconfinement increases for very high values of the rotating wall potential, which is what we expect. In our analysis, we stay in regions where the extent of crystal is much smaller than the separatrix radius.

We also need to exclude unstable equilibrium configurations (which are indicated by the nonpositivity of the eigenvalues $\lambda^{z\nu}$ of the stiffness matrices of the axial vibrations). We examine this in detail below, when we discuss our results on the axial phonon modes.


To find the equilibrium configurations, we need to minimize the Hamiltonian of the crystal in the rotating frame, which boils down to finding the best minimum of the transformed potential function of Eq.~(\ref{eq: potential_dimensionless}) near a triangular lattice. We only concentrate on the solution we obtain starting from the closed-shell construction. In this case, we start with a seed lattice where we arrange ions in closed, triangular shells while also respecting the triangular lattice symmetry. The shell is closed if we can put all the ions in these complete  shells. We relate the number of shells $S$ to the number of total ions in the crystal $N$ as
\begin{equation}
S = \left[ \sqrt{\frac{2(N-1)}{3} +\frac{1}{4}} - \frac{1}{2}\right].
\end{equation}
If we cannot put the $N$ ions in an integer number of shells, we put the outermost ions in an incomplete triangular ring according to the minimal potential energy at each of the outer ring sites. For the purpose of our discussion here, we pick $N=85$ and hence $S=7$.  We arrive at the minima guaranteed under such a consideration of the seed lattice using a trust-region algorithm of the MATLAB Optimization Toolbox. The minimization procedure requires us to specify a locally calculated gradient of the potential, which can be input analytically by taking derivatives of the potential. The procedure iterates the minimization steps until the local minimum is found.

\begin{figure}[ht]

\centering
\hspace{-7mm}
\begin{minipage}[b]{0.20\linewidth}
\includegraphics[scale=0.145]{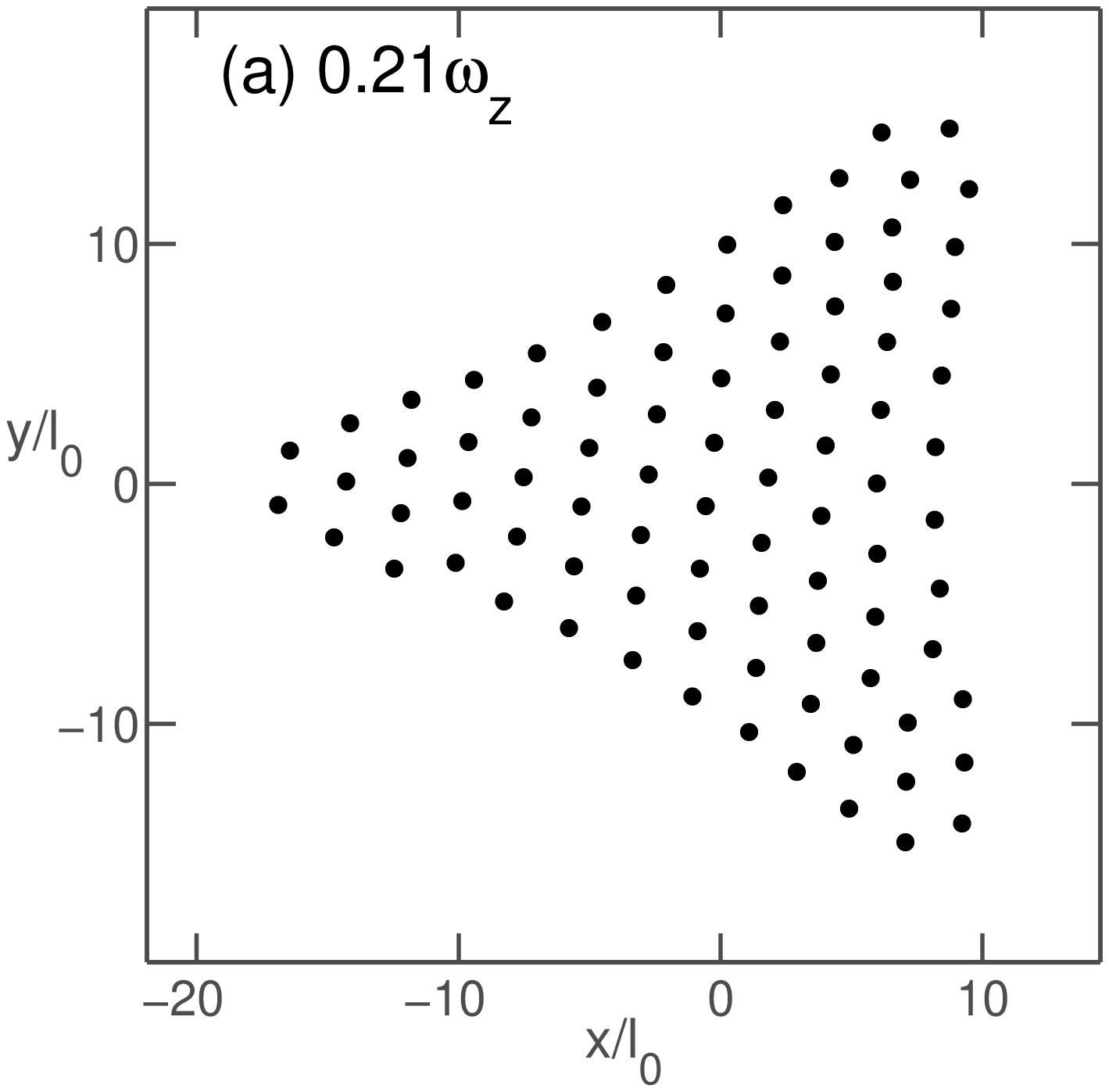}
\end{minipage}
\hspace{3mm}
\begin{minipage}[b]{0.20\linewidth}
\includegraphics[scale=0.145]{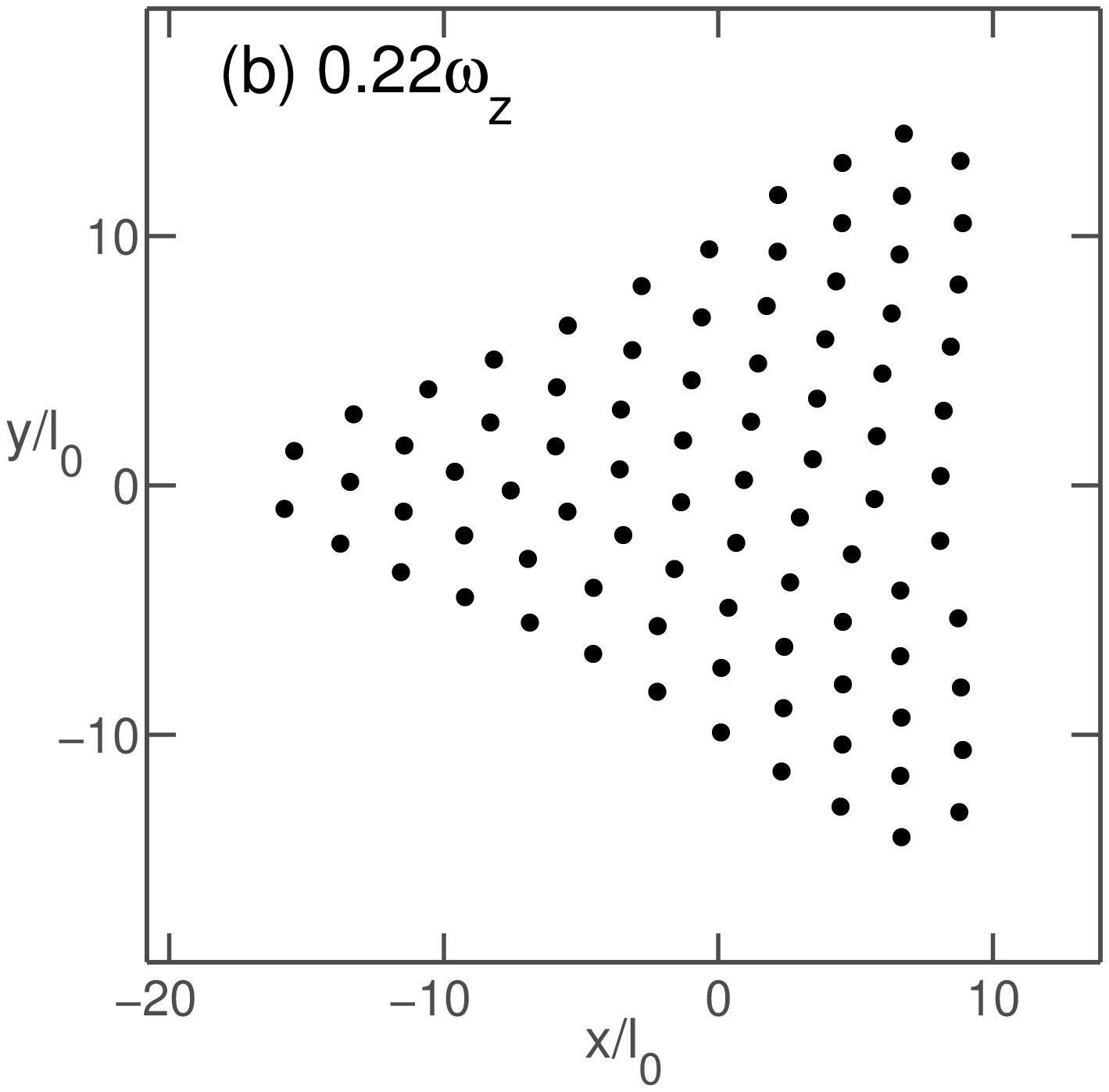}
\end{minipage}
\hspace{3mm}
\begin{minipage}[b]{0.20\linewidth}
\includegraphics[scale=0.145]{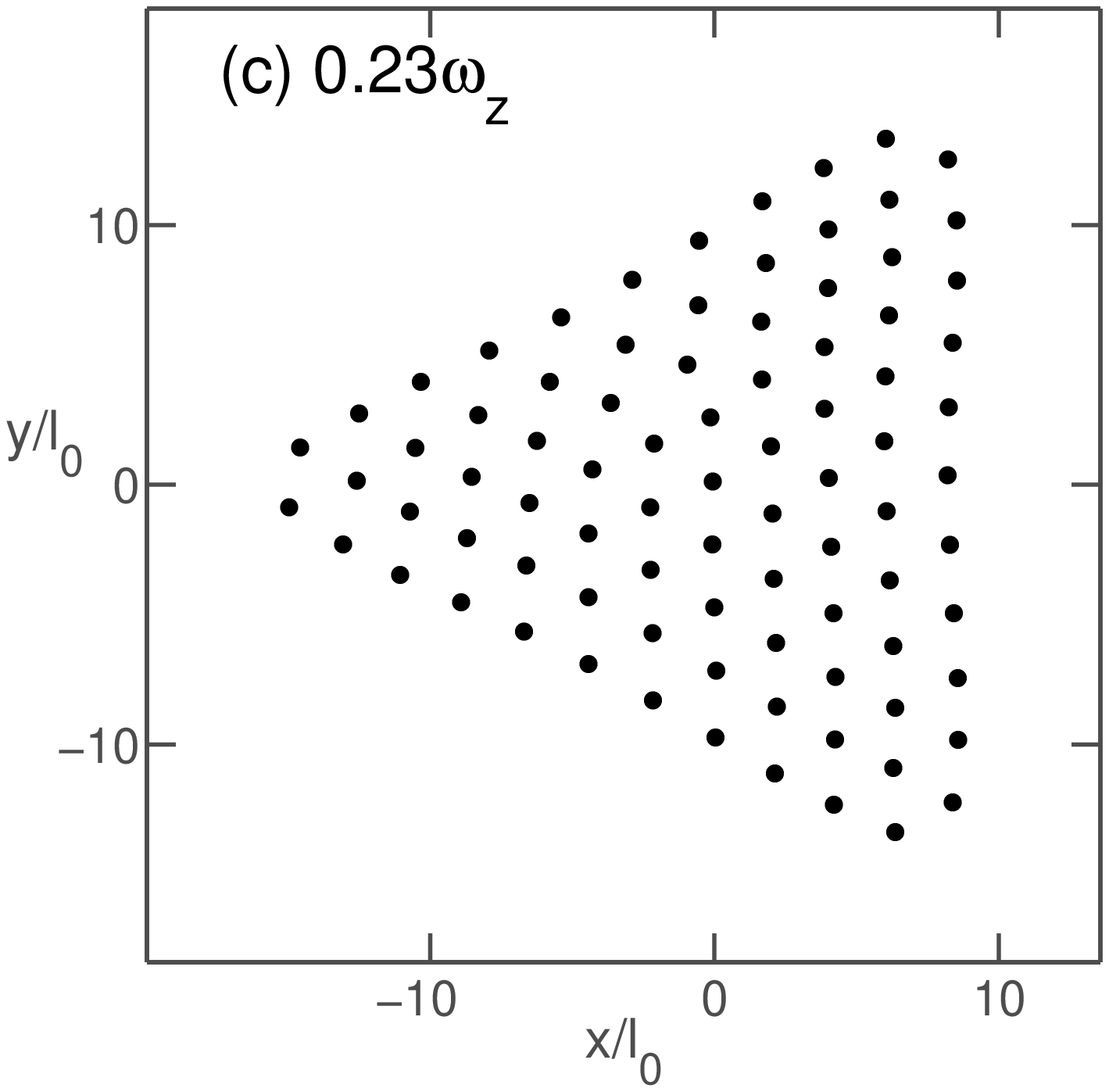}
\end{minipage}
\hspace{3mm}
\begin{minipage}[b]{0.20\linewidth}
\includegraphics[scale=0.145]{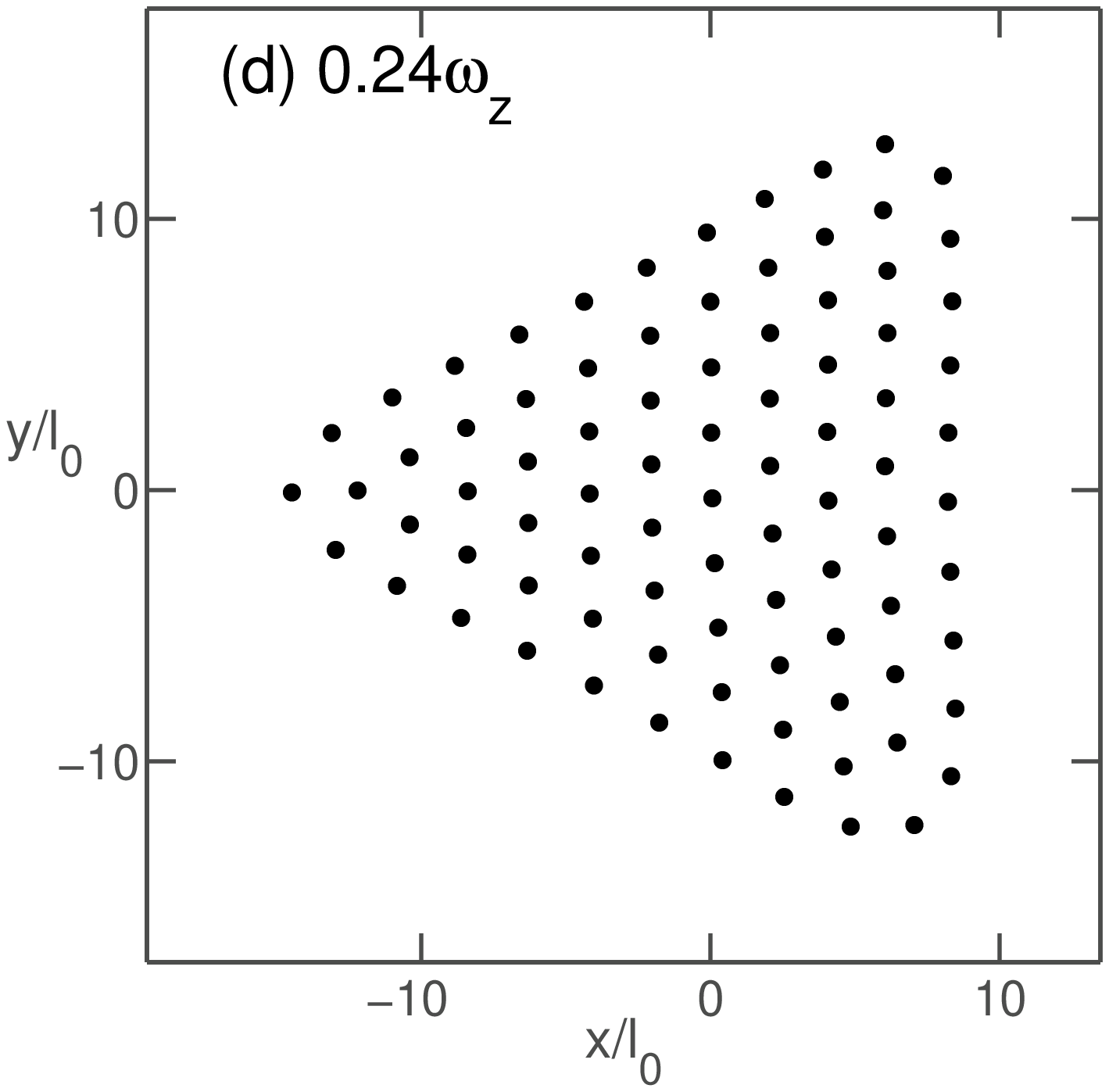}
\end{minipage}

\centering
\hspace{-7mm}
\begin{minipage}[b]{0.20\linewidth}
\includegraphics[scale=0.145]{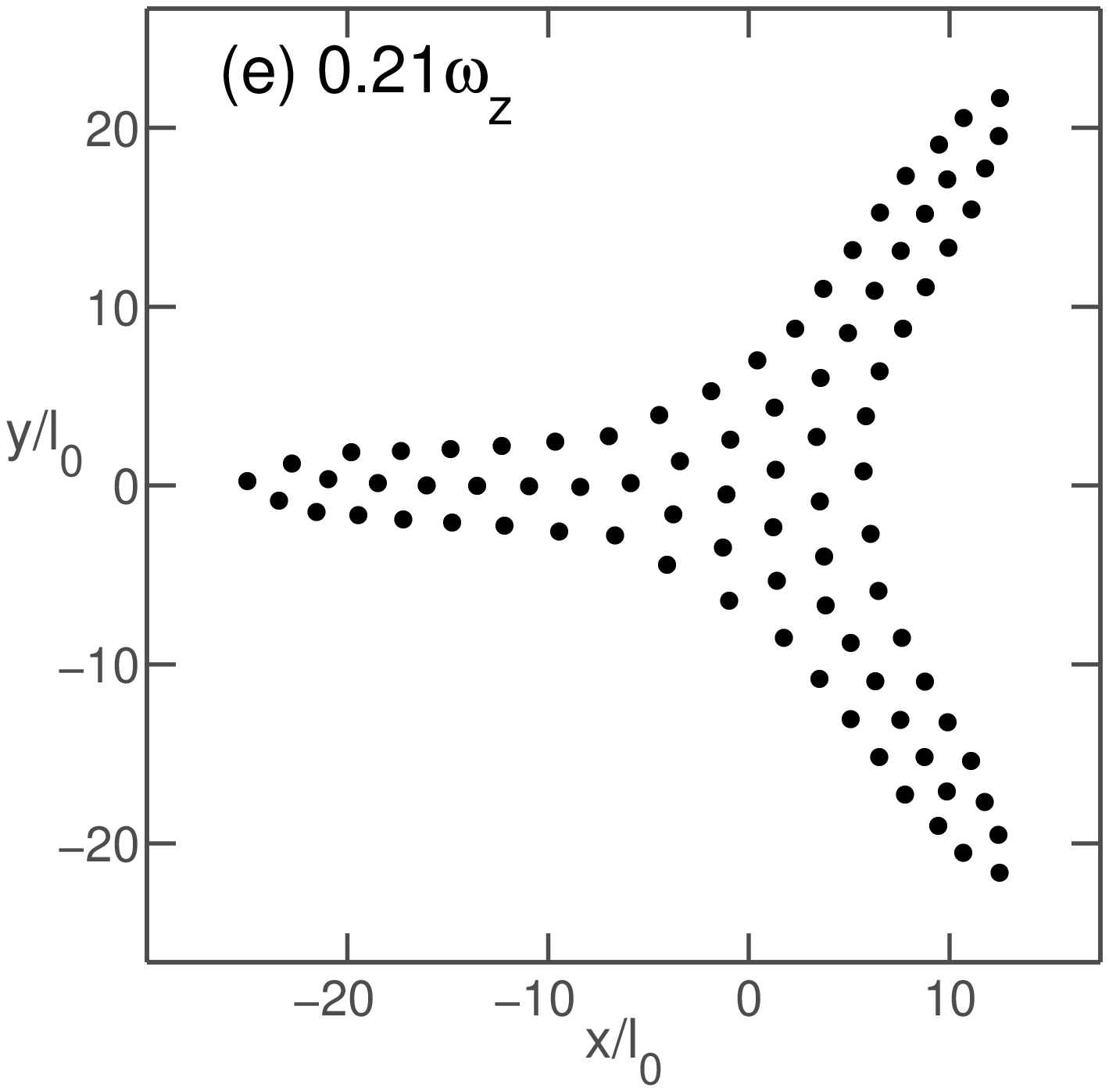}
\end{minipage}
\hspace{3mm}
\begin{minipage}[b]{0.20\linewidth}
\includegraphics[scale=0.145]{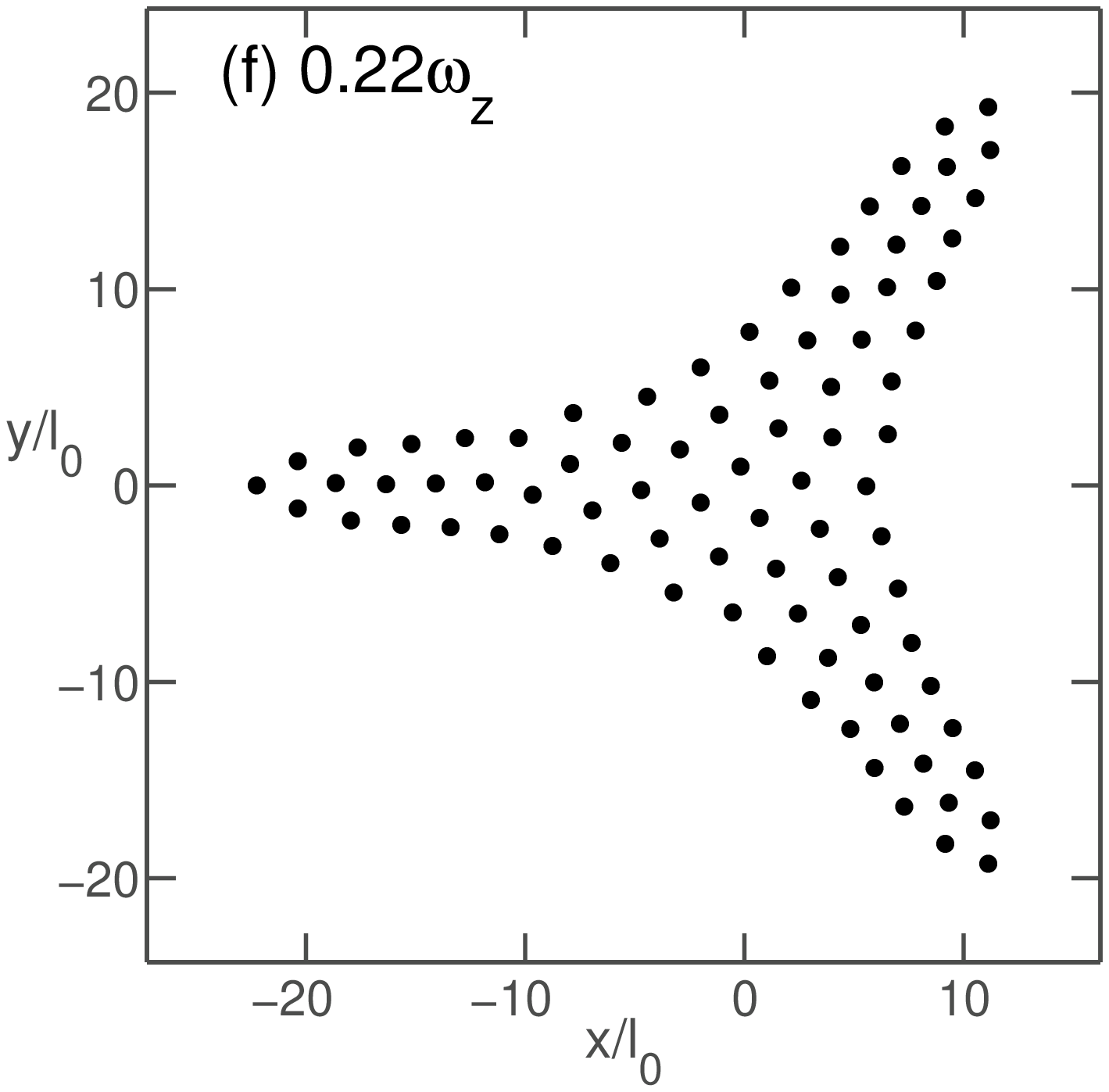}
\end{minipage}
\hspace{3mm}
\begin{minipage}[b]{0.20\linewidth}
\includegraphics[scale=0.145]{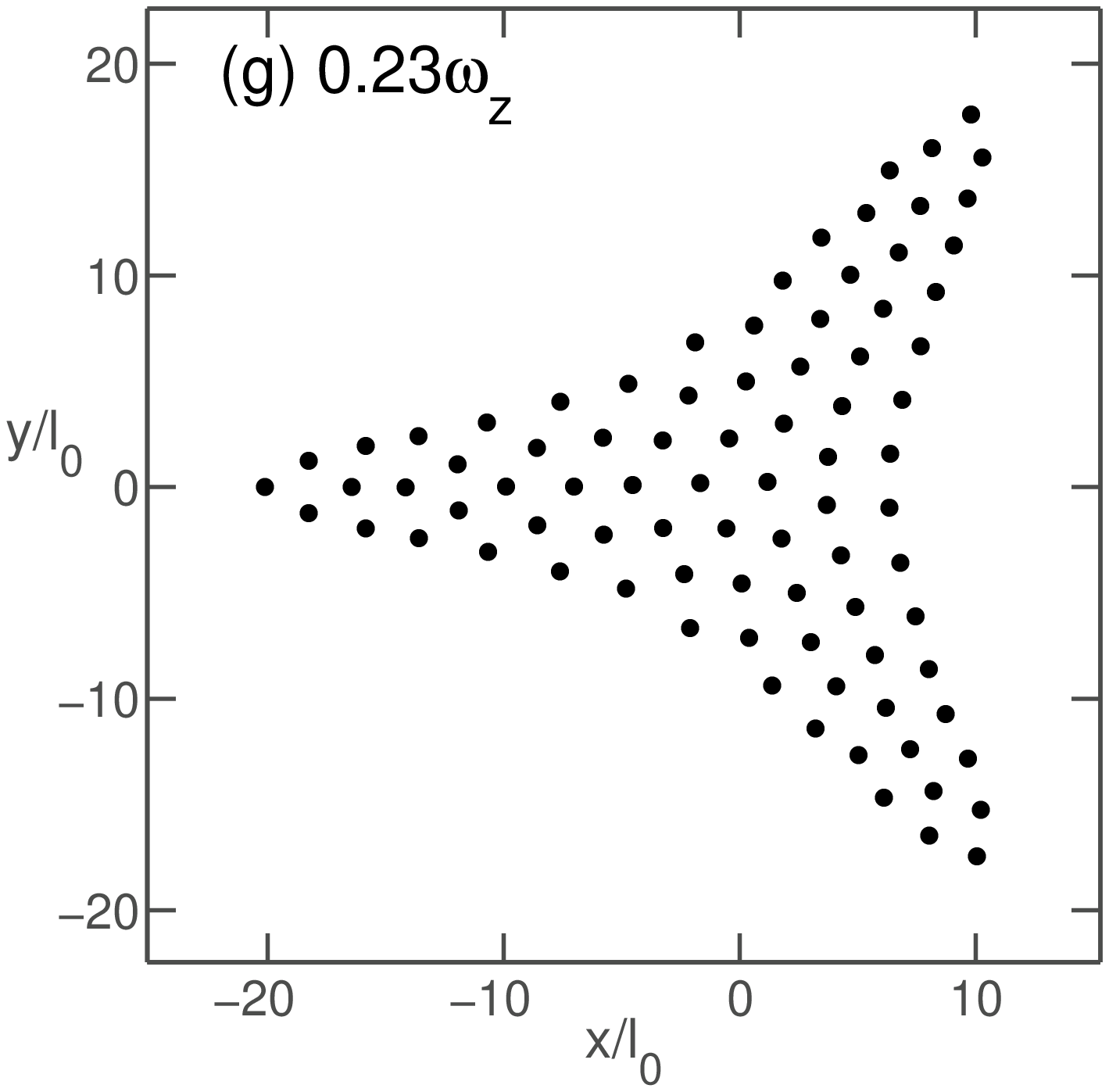}
\end{minipage}
\hspace{3mm}
\begin{minipage}[b]{0.20\linewidth}
\includegraphics[scale=0.145]{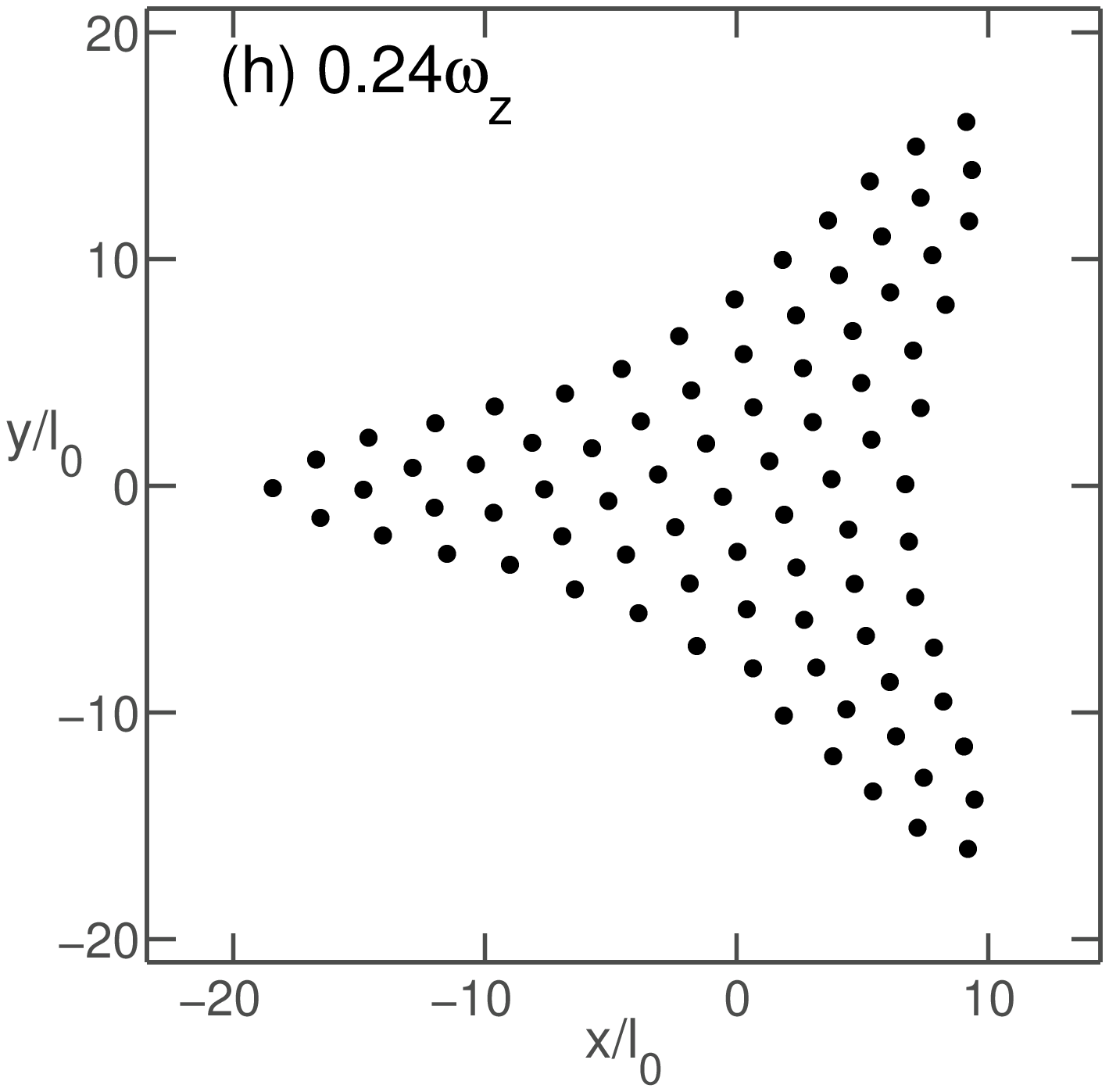}
\end{minipage}

\caption{Equilibrium structures found for varying $\omega_{eff}$ and $V_{W}$. Panels (a)-(d) show the succession of crystal structures obtained by increasing $\omega_{eff}$ with $V_{W}=0.0025\omega_{z}^{2}$. Panels (e)-(h) present the structures for higher $V_{W}=0.0040\omega_{z}^{2}$\, for the same corresponding values of $\omega_{eff}$. 
\label{fig: equilib}}

\end{figure}

The equilibrium configurations we obtain from such a procedure behave as we might expect (see Fig.~\ref{fig: equilib}). We obtain structures that form a nearly perfect triangular lattice close to the center, and smoothly transition to the shape of the contour lines of the effective potential as we move radially outward to the edges. The edge effects cause the interionic distances to change as we move outwards. Our strategy to counter these effects are two-fold, as discussed earlier: (i) We introduce a weak anharmonic term whose strength is characterized by $C_{4}$ and (ii) we match the symmetry of the rotating wall with that of the condensed crystal. The first effect has been incorporated and fixed at a particular value, as discussed earlier. However, we keep the strength of the rotating wall variable as it is only for a certain range of values of the strength of the rotating wall potential that the contour lines of the effective potential are triangular, and hence become least likely to cause edge distortions of the ionic crystal.

This is clearly seen in the Fig.~\ref{fig: equilib}, where we show the progression of these structures for different values of the effective radial trapping strength $\omega_{eff}$, for $V_{W}^{l}=0.0025$ and $V_{W}^{h}=0.0040$. For the low-strength rotating wall, the structures (a)-(d) are nearly triangular and uniform, whereas the higher value of the rotating wall strength corresponds to the more distorted structures of (e)-(h). For a fixed number of ions and fixed $C_4$, we see the destabilizing effect of the decrease in separatrix radius with increasing rotating wall strength clearly on these crystal structures. We note here that in the limit of vanishing wall strength or high effective radial trapping frequency, the structures become more isotropic, with uniformly decreasing nearest-neighbor distances as we move towards the edges. The other extreme (very small radial trapping frequency or high strength of rotating wall), the equilibrium configuration we obtain from a closed-shell construction shows that the ions are reduced to (three) pockets of stability and the structure is no longer closed. A detailed normal-mode analysis (see below) shows that these structures are in unstable equilibrium, and hence we can discard them. Another important observation is that a given progression of structures (for differing values of the radial trapping strength and increasing values of $V_W$)  displays similar structures to those found at smaller trapping strengths, for higher values of the rotating wall strength. This fact will be important to arrive at structures that show the maximum uniformity and also exhibit a stable equilibrium. 

Note that in these calculations, we fix the wall potential and then vary $\omega_{eff}$. In doing so, we find that we are limited by how many ions we can hold in stable equilibrium.  Because the
radial deconfinement decreases as the rotational frequency $\omega_{eff}$ decreases, the effect of the rotating wall will become stronger if it also remains fixed. We have done this here to reduce the parameter space we explore. But, if one wants to examine larger crystals, then one needs to 
carefully tune the rotating wall strength as the rotational frequency is changed, as well as the 
strength of the quartic potential, to be able to continue to maintain stable equilibrium. These issues are discussed in Ref.~\cite{dubin}

\begin{figure}[ht]
\centering
\includegraphics[scale=0.45]{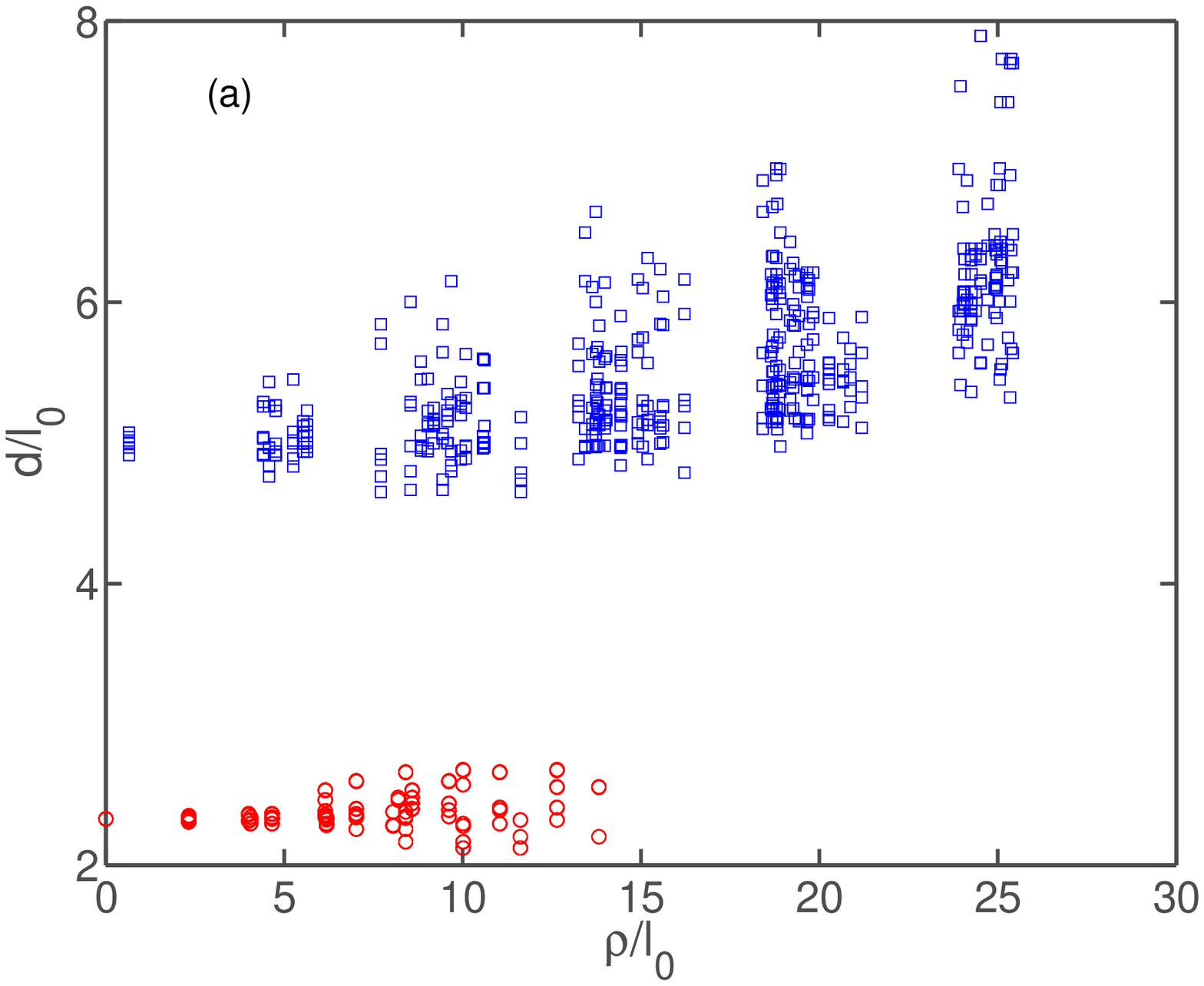}
\centering
\includegraphics[scale=0.26]{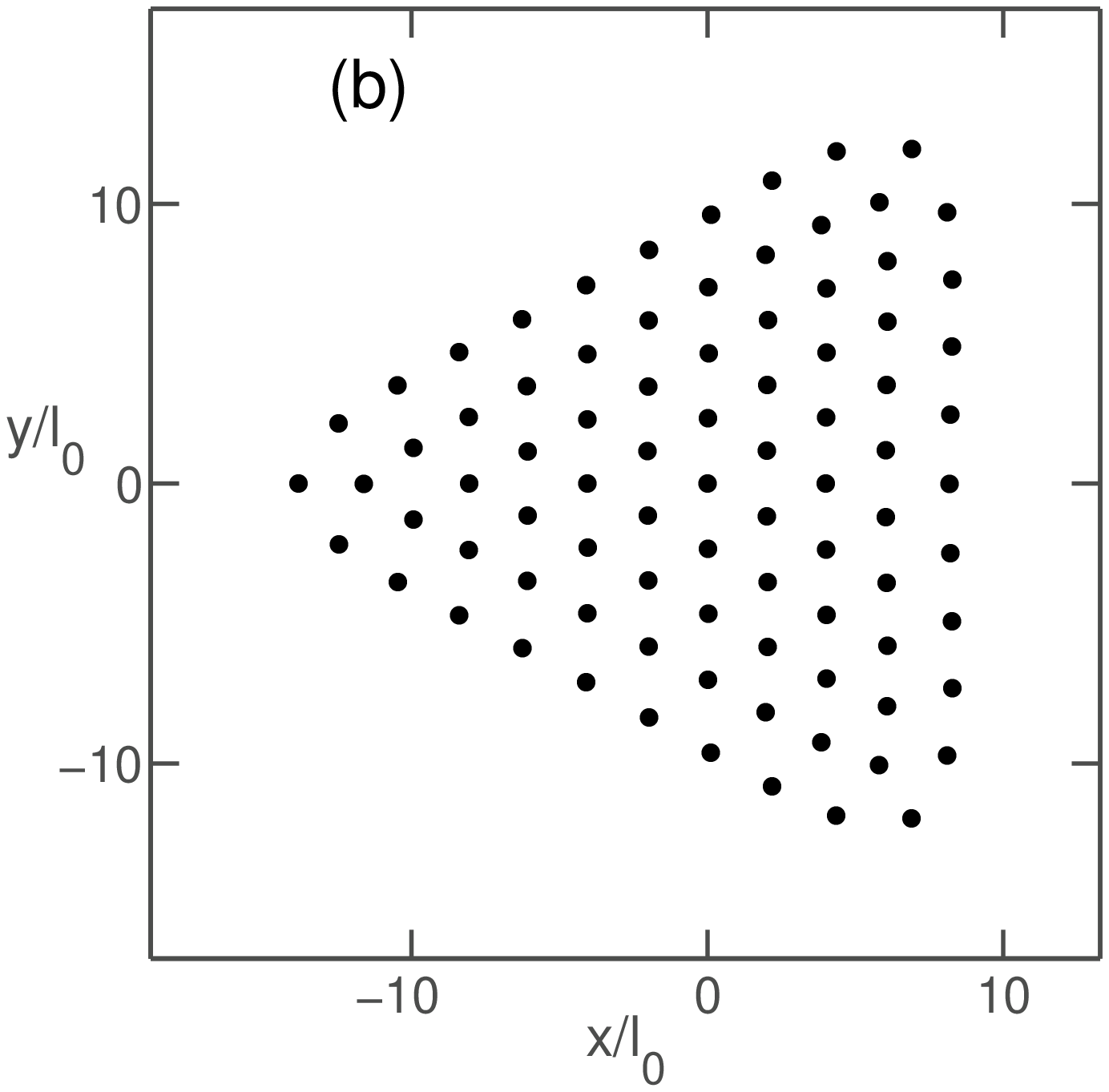}\includegraphics[scale=0.26]{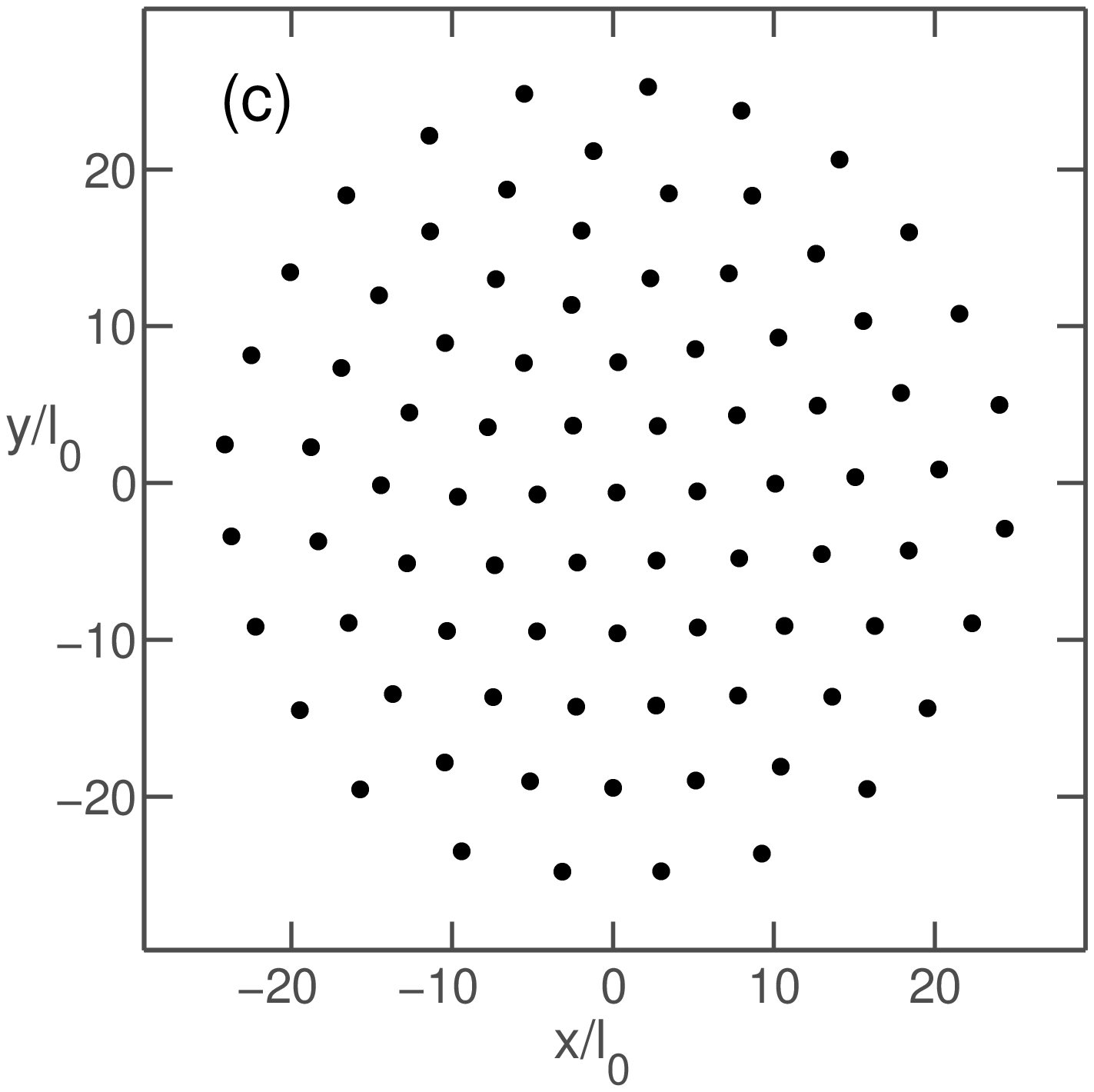}
\caption{(Color online.) Nearest-neighbor distances $d$ versus the distance of the ion from the trap symmetry axis $\rho$, where all distances have been normalized against the length scale $l_0$. The red circles plot this for the $l=3$  triangular wall with an additional anharmonic potential for $V_{W} = 0.0025 \omega_{z}^{2}$ and $\omega_{eff} = 0.25\omega_{z}$ [equilibrium positions shown in panel (b)], while the blue squares represent this for the original $l=2$ rotating wall with just a harmonic potential for $\omega_{eff} = 0.06\omega_{z}$ [equilibrium positions shown in panel (c)]. The parameters for the former trap are chosen so as to minimize the variance in nearest-neighbor distances, while a moderate wall potential is chosen for the latter for a valid comparison. The variance in the triangular wall lattice (b)  is much smaller than in the quadrupole wall lattice (c).  \label{fig: distance}}
\end{figure}

The equilibrium structures we obtain are markedly uniform, and this is born out in Fig.~\ref{fig: distance} where we plot the distance to the nearest neighbor for each ion in the crystal as a function of the central ions' distance from the trap symmetry axis. These results are for the most uniform stable structures we could obtain for both the $l=2$ and $l=3$ rotating walls. The nearest-neighbor distances have been calculated based on the Delaunay triangulation algorithm. Each point in the figure represents the distance of the ion in question to an ion in the first nearest-neighbor shell. We focus on the first circle of nearest neighbors only. 
The values of the relevant parameters, for both the $l=2$ (quadrupole) and the $l=3$ (triangular) wall crystals, are indicated in the caption to Fig.~\ref{fig: distance}. The larger spread of values for the quadrupole wall tells us that, on average, the triangular wall crystal is indeed more uniform spatially than the quadrupole wall crystal. Note that there are the same number of blue squares and red circles in Fig.~\ref{fig: distance}; the uniformity of the triangular lattice has many of the symbols overlap.

\begin{figure}[ht]
\centerline{
\includegraphics[scale=0.39]{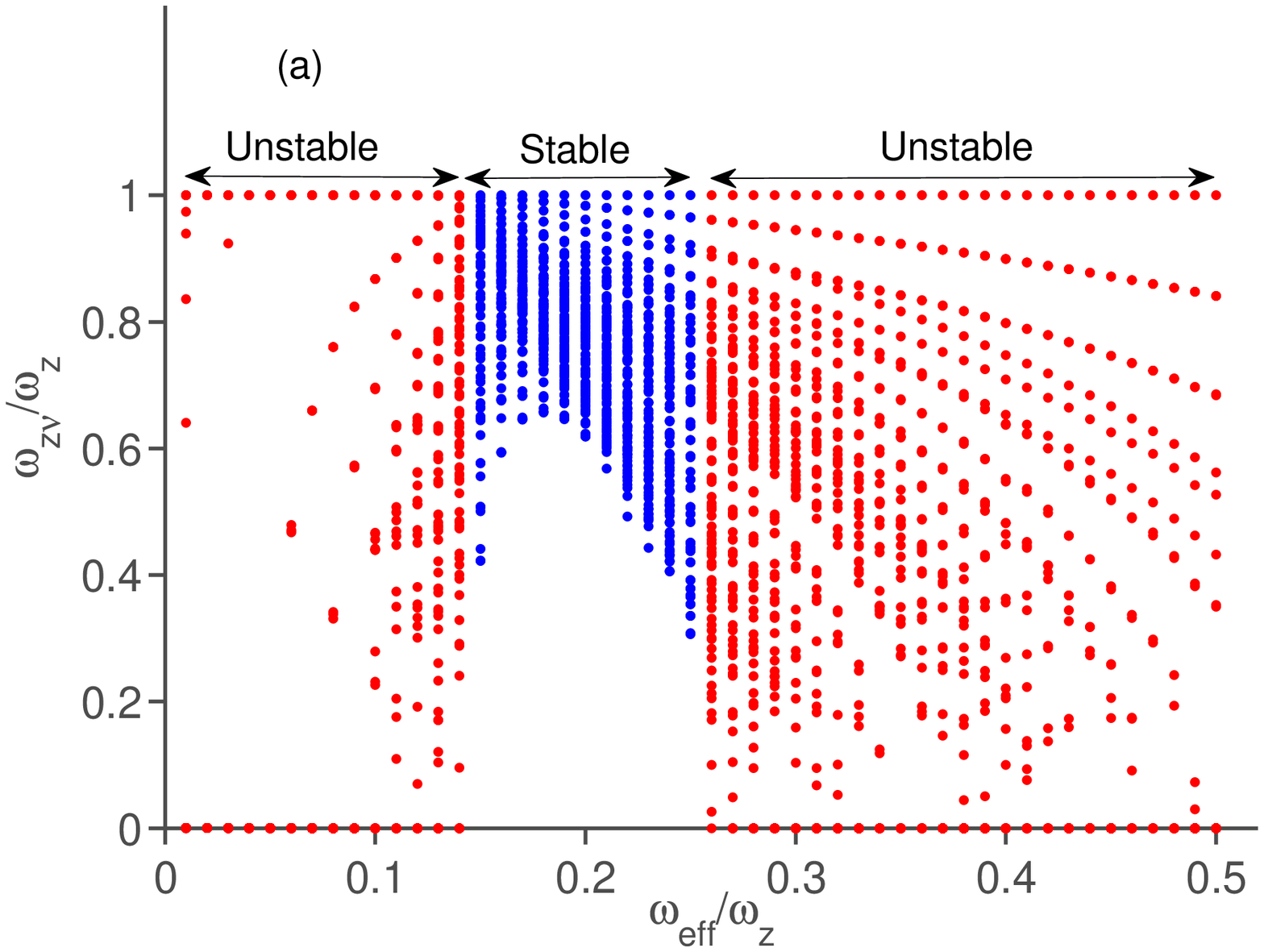}}
\centerline{
\includegraphics[scale=0.39]{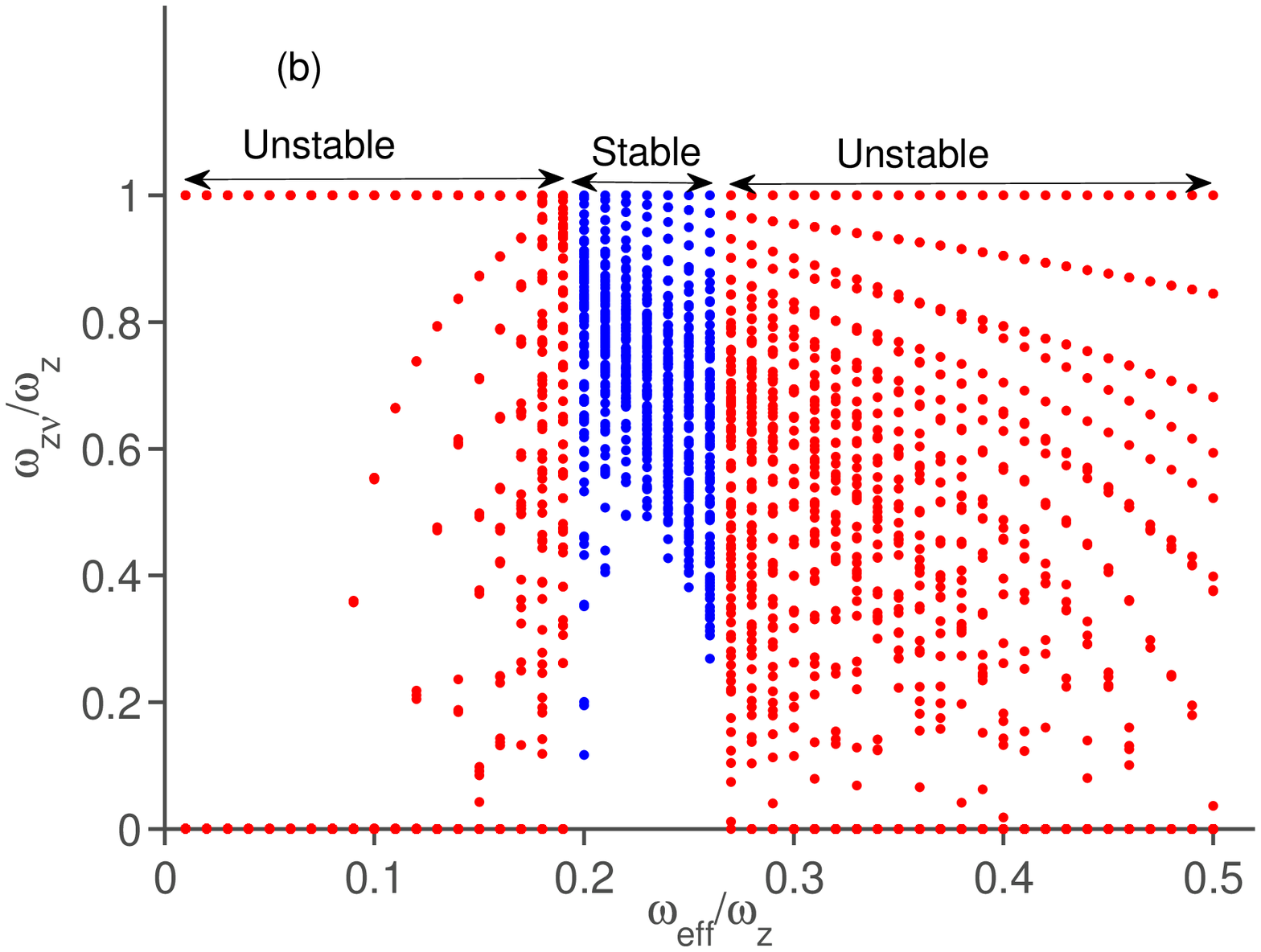}}

\caption{(Color online.) Variation of the axial normal mode frequencies (in units of $\omega_{z}$) with $\omega_{eff}$, for two wall strengths: (a) $V_{W}$=0.0025$\omega_{z}^{2}$ and (b) $V_{W}$=0.0040$\omega_{z}^{2}$. \label{fig: normal_modes} }

\end{figure}

We next discuss the features of the normal modes of small oscillations of the ions. We first examine the positivity of the eigenvalue spectrum of the stiffness matrix of axial vibrations $\mathbf{K^{zz}}$ as a function of effective trapping strength, for high and low rotating wall potential strengths, in Fig.~\ref{fig: normal_modes}. We see that the eigenvalue spectrum is real and positive for only a ``band" of values of $\omega_{eff}$ (indicated in blue in the figure), and this band shrinks for the higher amplitude rotating wall potential. The positivity of all the eigenvalues indicates stability of the corresponding structures.

This behavior is fundamentally different from that seen in the usual Penning trap crystals where we do not see a lower bound to the strength of the radial trapping, characterized by $\omega_{eff}$, where the structures become unstable, although there is an upper limit. Hence, we can only talk of stable structures for certain ranges of $\omega_{eff}$ for the $l=3$ rotating wall (with an additional quartic potential), and this range gets narrower for increasing wall strength.  A similar effect is seen if the number of ions in the trap is increased, in that the band of stability shrinks as we increase the number of ions in the trap. The value of $N=85$ was the maximum number of ions we found that could be trapped with reasonably large bands of stability for the particular ranges of parameters that
we chose. Note that more ions can be trapped by carefully choosing the rotation frequency $\omega_{eff}$, the rotating wall potential amplitude, and the strength of the quartic potential, but we do not discuss these cases in detail here.

\begin{figure}[ht]
\centering
\includegraphics[scale=0.45]{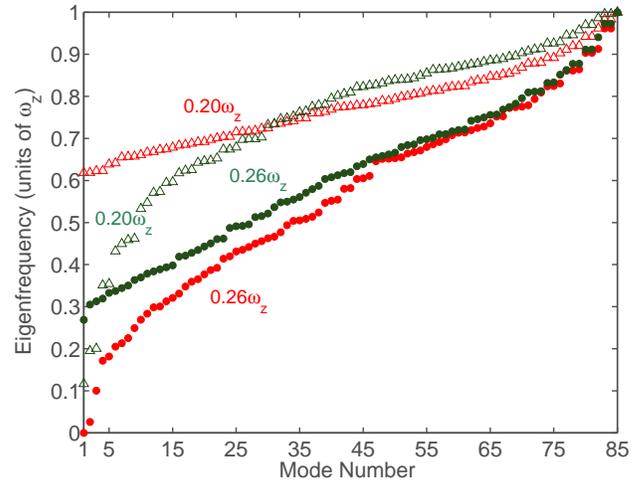}
\caption{(Color online.) Eigenfrequencies of the axial phonon modes. The green and red symbols represent data for a strong rotating wall, $V_{W} = 0.0040\omega_{z}^{2}$, and a weak wall $V_{W} = 0.0025\omega_{z}^{2}$, respectively. The values of the corresponding effective trapping frequency $\omega_{eff}$ are indicated by the labels near each curve (hollow symbols, $\omega_{eff}=0.20\omega_z$ and solid symbols, $\omega_{eff}=0.26\omega_z$). Note how the strong rotating wall and high trapping frequency case is nearly unstable.\label{fig: normal_modes2}}
\end{figure}

Next, we plot the numerically obtained eigenfrequencies against their mode numbers, for crystal structures corresponding to two values of (stable) $\omega_{eff}$, for both high and low rotating wall potentials. Roughly, we can see a trend similar to that of the quadrupole-wall crystal normal modes, where the primary dependence is on $\omega_{eff}$ and not on $V_W$. However, there are important distinctions to be made. The ``band" structure of the eigenvalue spectrum causes the structure of these curves to change, and quite significantly at that, when we vary the rotating wall from low to high strength. We see this in Fig.~\ref{fig: normal_modes}, where for $\omega_{eff}=0.20\omega_{z}$, the lower edge of the instability of the band shown in Fig.\ref{fig: normal_modes} shifts to the right when the strength of the rotating wall is increased, and this causes the eigenfrequencies for the higher wall strength to drop abruptly to values very close to zero. In this fashion, we see that the dependence on $\omega_{eff}$ is now superimposed on a dependence on the strength of the rotating wall due to a $V_{W}$-variable bandwidth.

The highest axial mode has a universal eigenfrequency equal to the angular frequency of the trapping strength, $\omega_{z}$ for all values of $\omega_{eff}$ and $V_{W}$, and we see that all the branches converge to this point. This behavior is identical to that of the quadrupole-wall rotating crystal. The corresponding eigenmode is the well-known center-of-mass mode, where all ions move equal displacements that are in-phase with each other. This is because the center-of-mass motion does not cost any additional Coulomb energy and all ions have the same axial trapping energy, and hence their motion is independent of the strength of the rotating wall potential applied in the crystal plane. Also, other axial phonon modes will have frequencies lower than the center-of-mass mode, as the average distance between the ions increases when the wave vector is nonzero and there is a reduction in energy due to the Coulomb repulsion. This is also why the axial eigenfrequency branches for higher $\omega_{eff}$ lie roughly lower in Fig.~\ref{fig: normal_modes2}. 

\begin{figure}[ht]
\hspace{-11mm}
\begin{minipage}[b]{0.16\linewidth}
\includegraphics[scale=0.145]{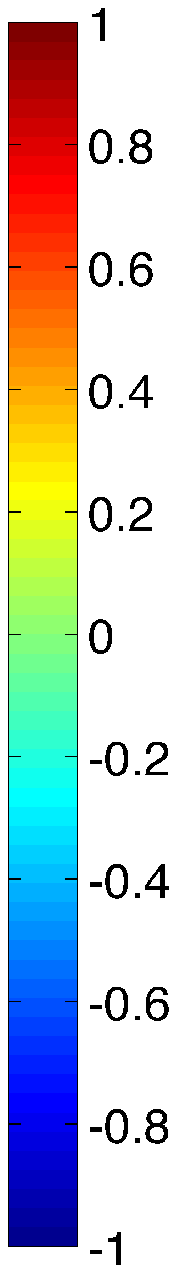}
\end{minipage}
\hspace{-5mm}
\begin{minipage}[b]{0.16\linewidth}
\includegraphics[scale=0.145]{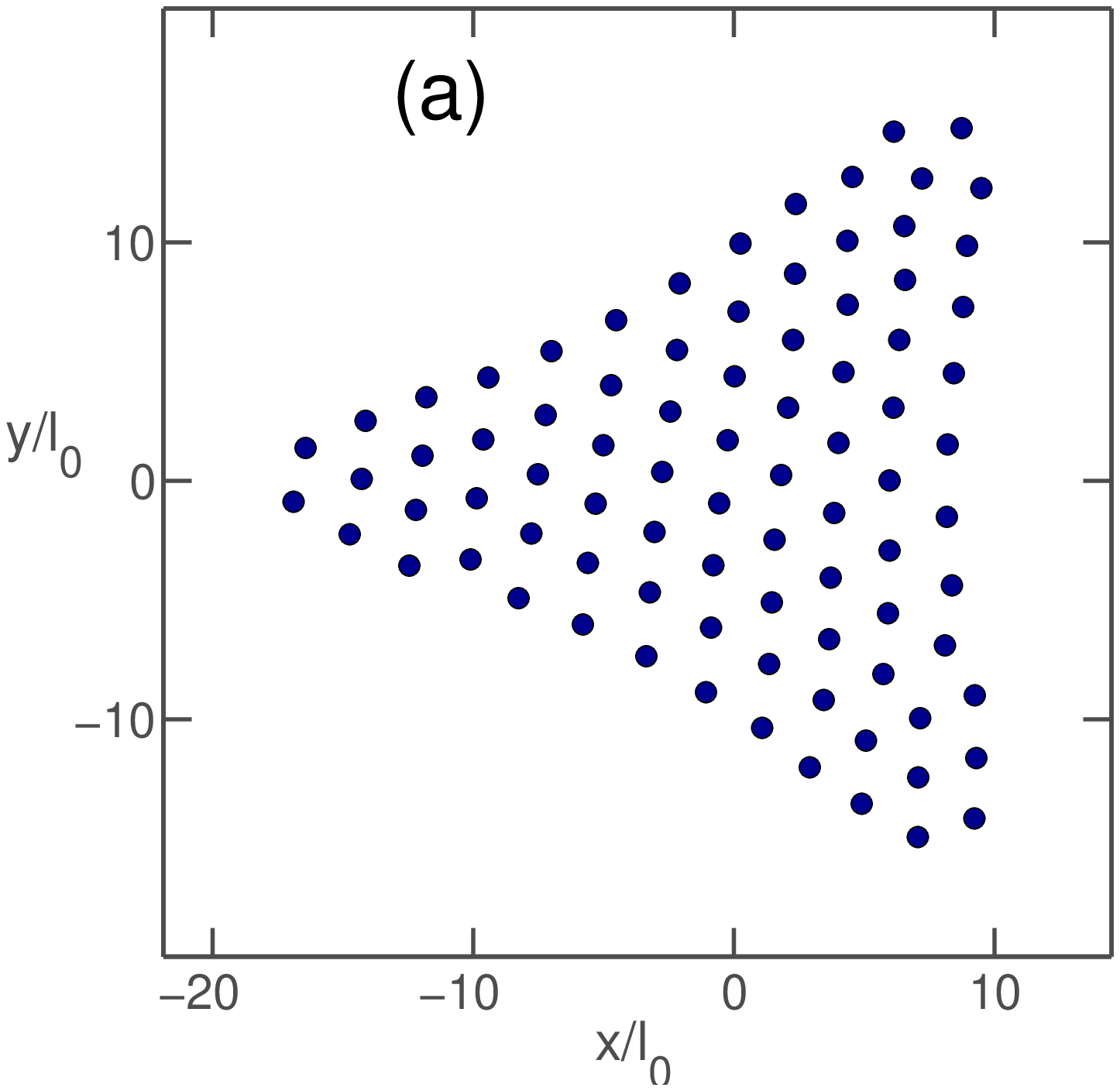}
\end{minipage}
\hspace{7mm}
\begin{minipage}[b]{0.16\linewidth}
\includegraphics[scale=0.145]{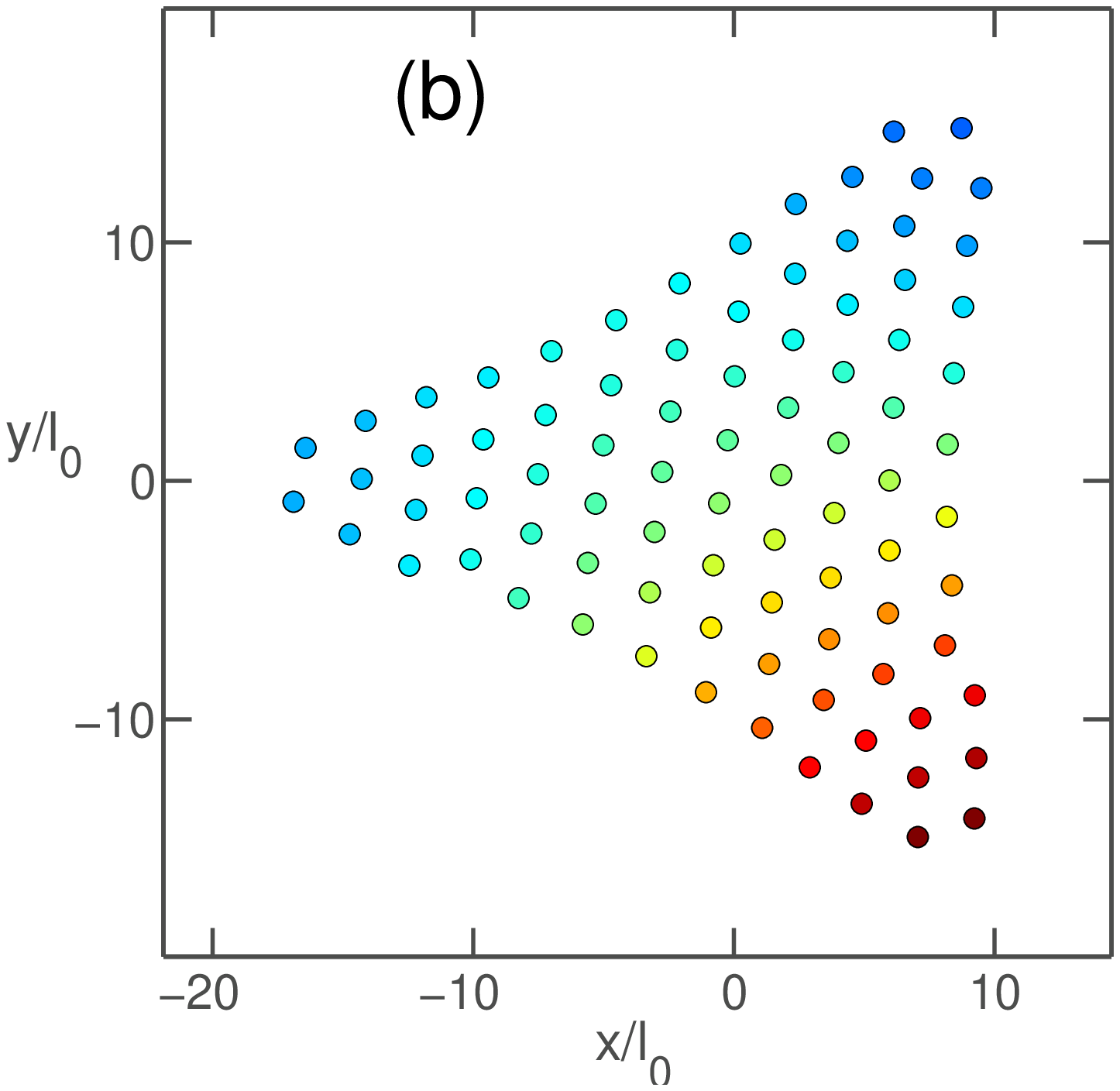}
\end{minipage}
\hspace{7mm}
\begin{minipage}[b]{0.16\linewidth}
\includegraphics[scale=0.145]{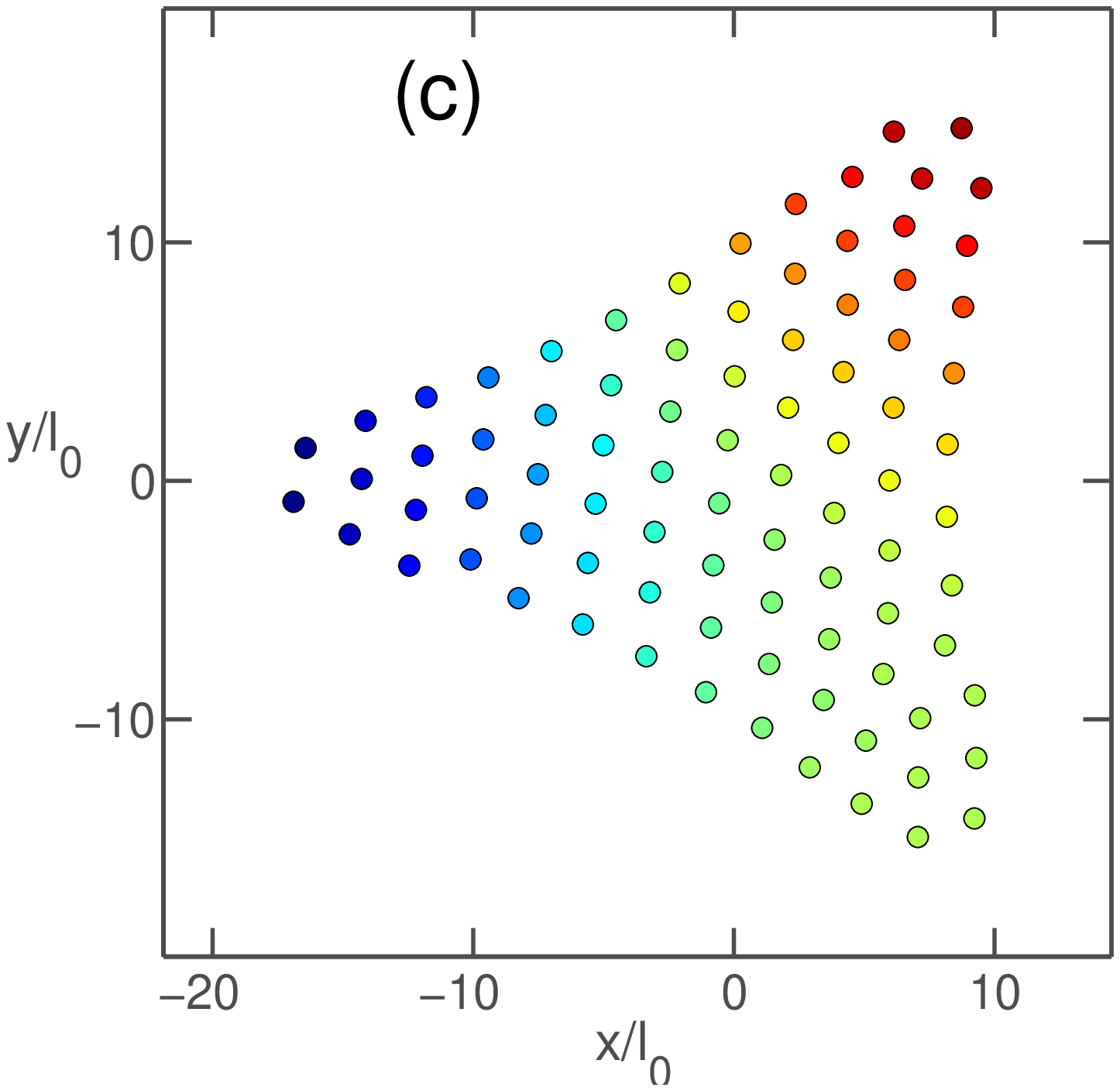}
\end{minipage}

\vspace{2mm}

\hspace{-11mm}
\begin{minipage}[b]{0.16\linewidth}
\includegraphics[scale=0.145]{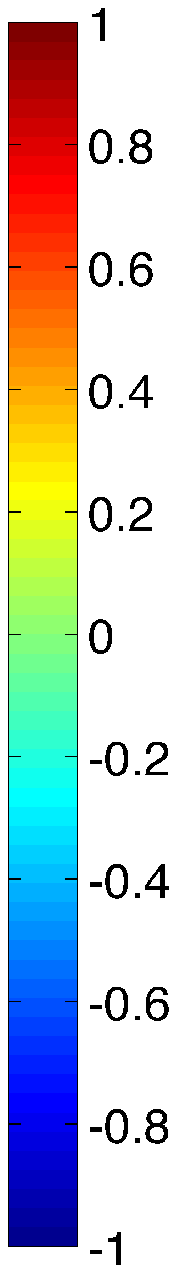}
\end{minipage}
\hspace{-5mm}
\begin{minipage}[b]{0.16\linewidth}
\includegraphics[scale=0.145]{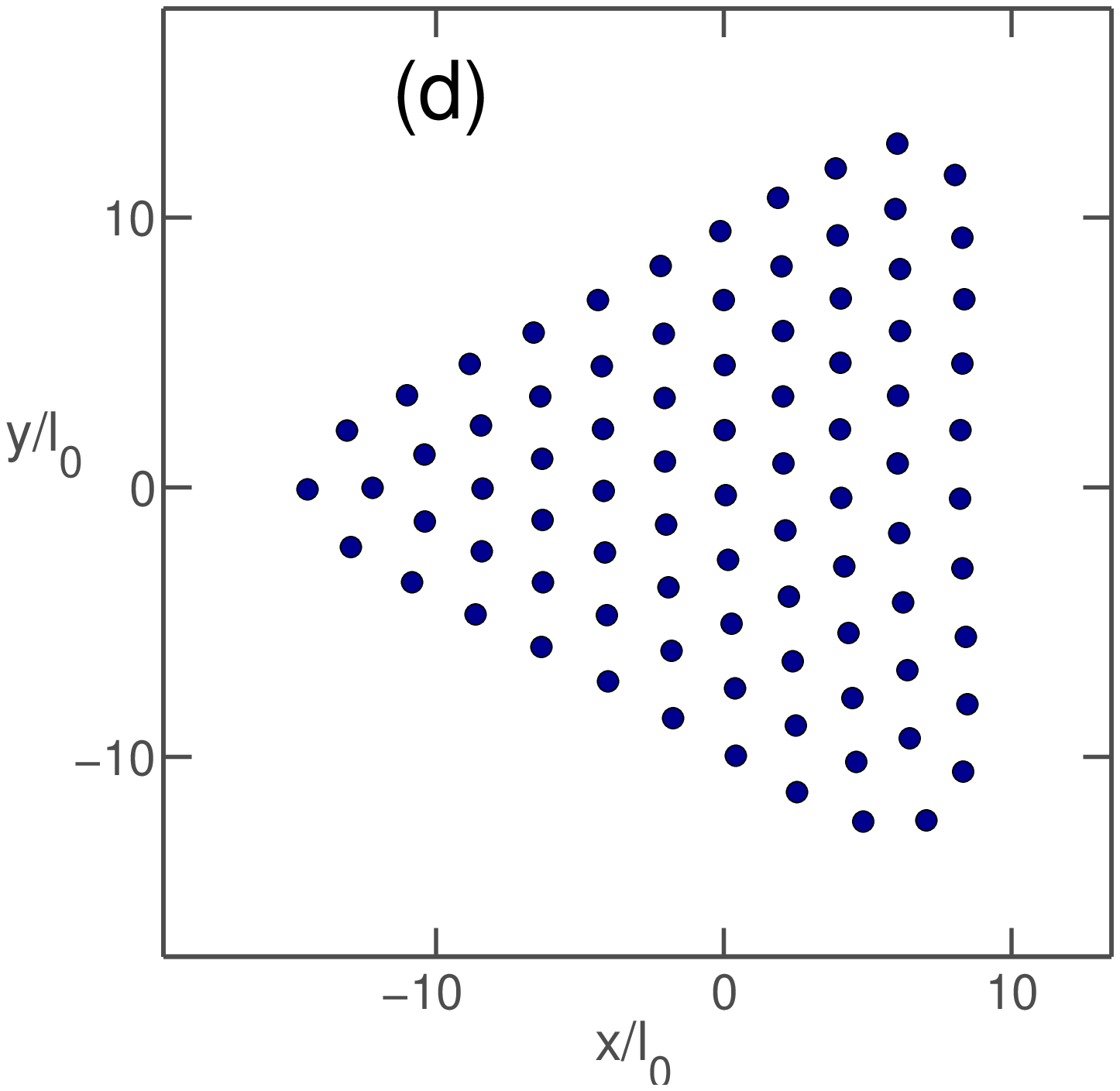}
\end{minipage}
\hspace{7mm}
\begin{minipage}[b]{0.16\linewidth}
\includegraphics[scale=0.145]{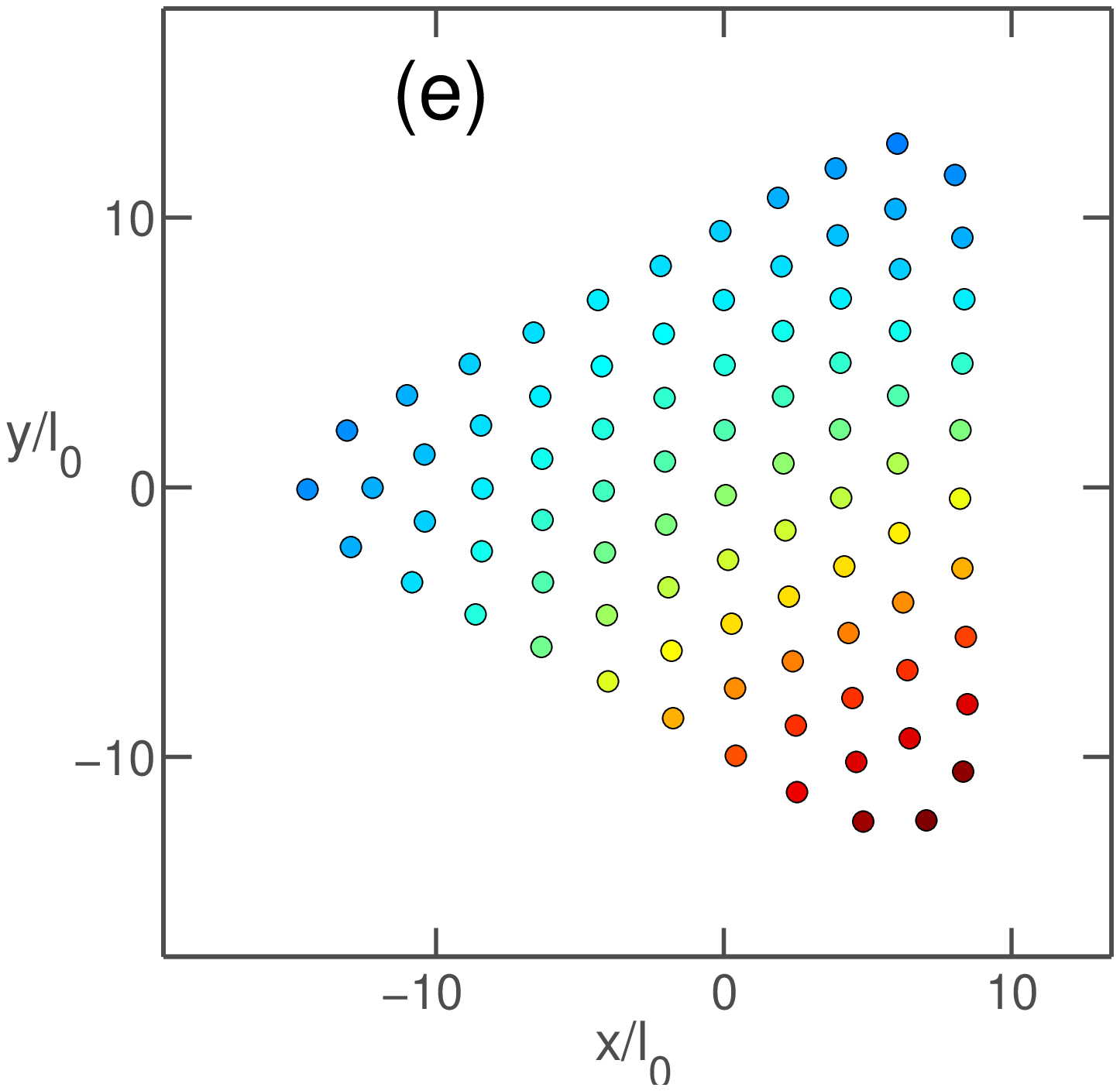}
\end{minipage}
\hspace{7mm}
\begin{minipage}[b]{0.16\linewidth}
\includegraphics[scale=0.145]{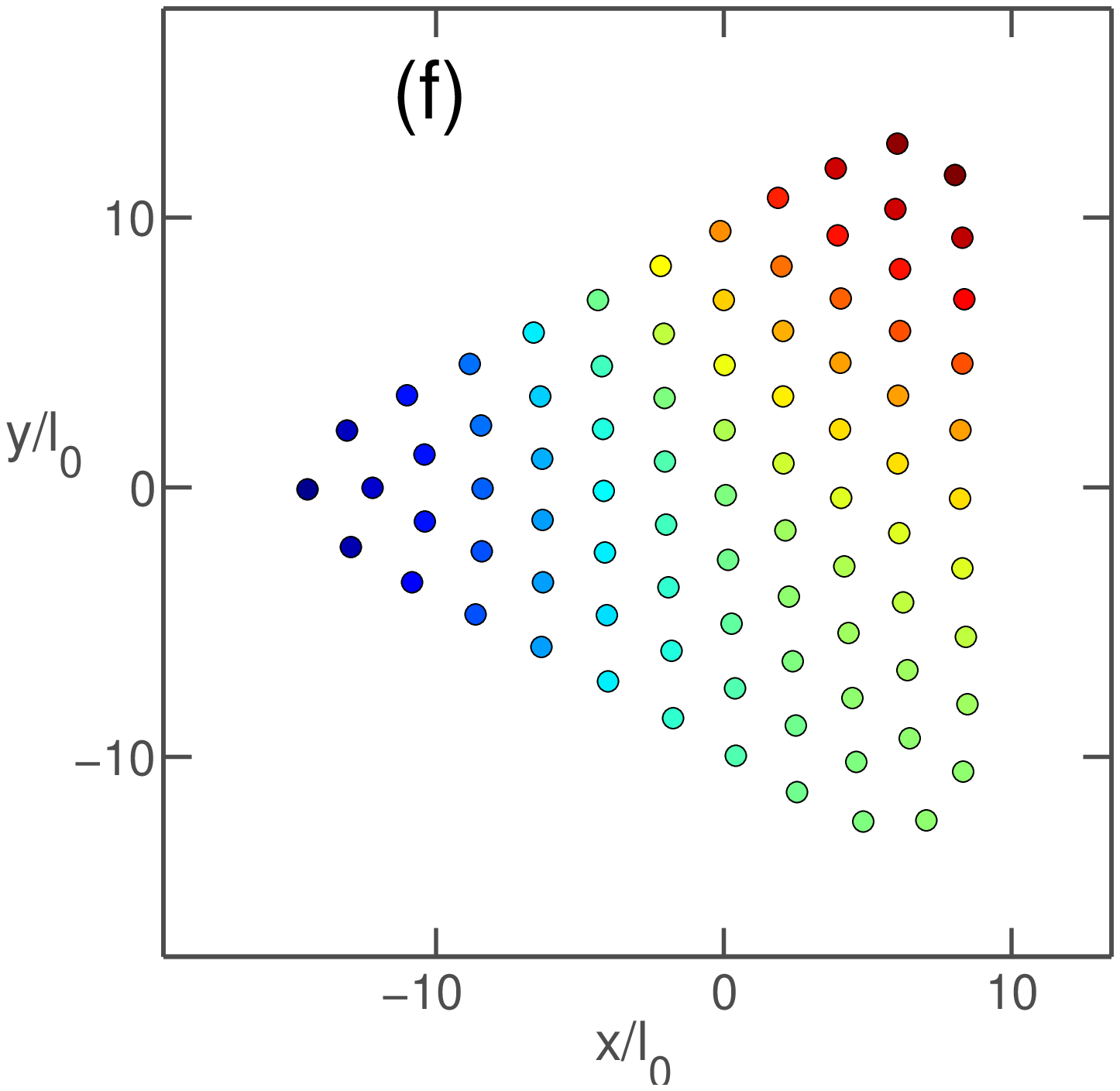}
\end{minipage}

\vspace{2mm}

\hspace{-11mm}
\begin{minipage}[b]{0.16\linewidth}
\includegraphics[scale=0.145]{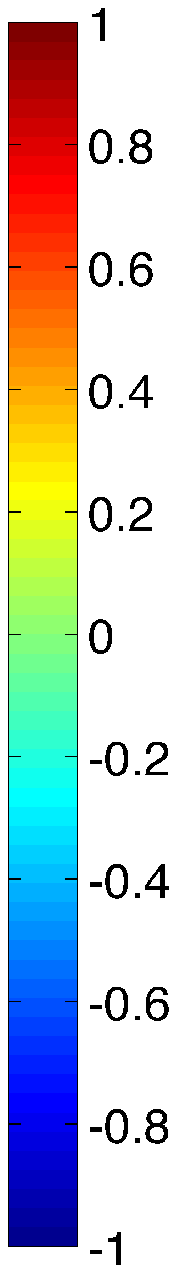}
\end{minipage}
\hspace{-5mm}
\begin{minipage}[b]{0.16\linewidth}
\includegraphics[scale=0.145]{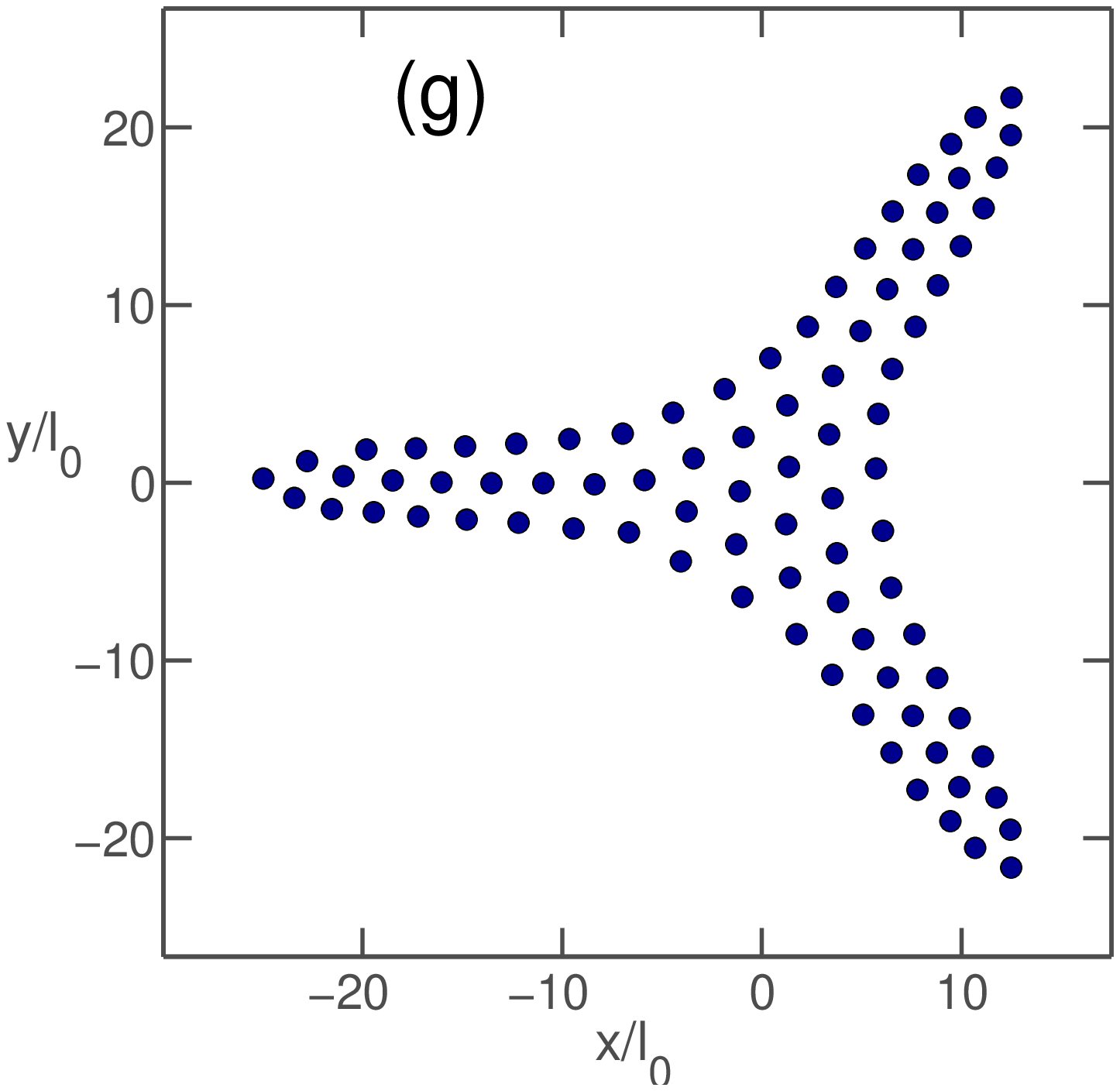}
\end{minipage}
\hspace{7mm}
\begin{minipage}[b]{0.16\linewidth}
\includegraphics[scale=0.145]{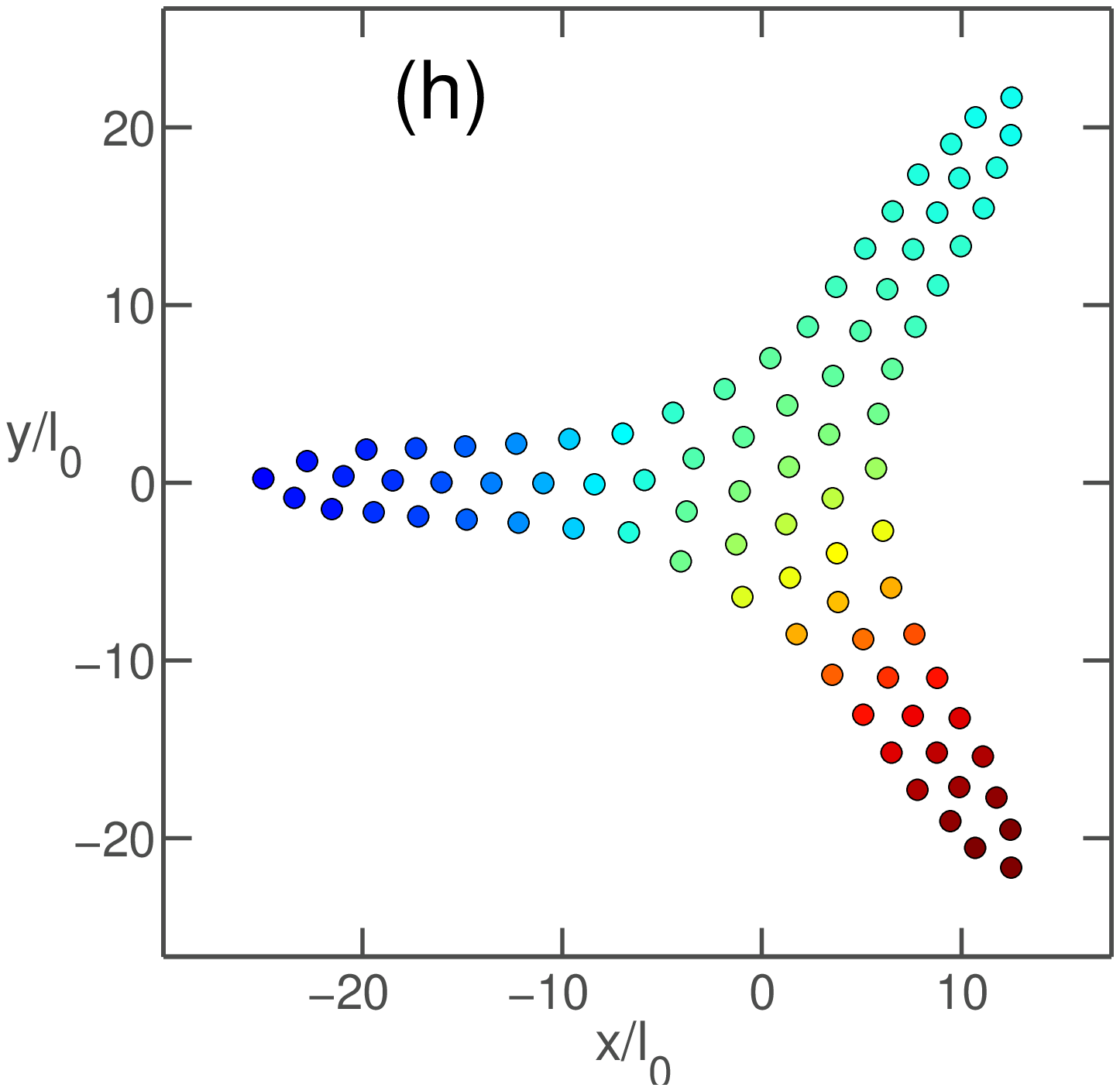}
\end{minipage}
\hspace{7mm}
\begin{minipage}[b]{0.16\linewidth}
\includegraphics[scale=0.145]{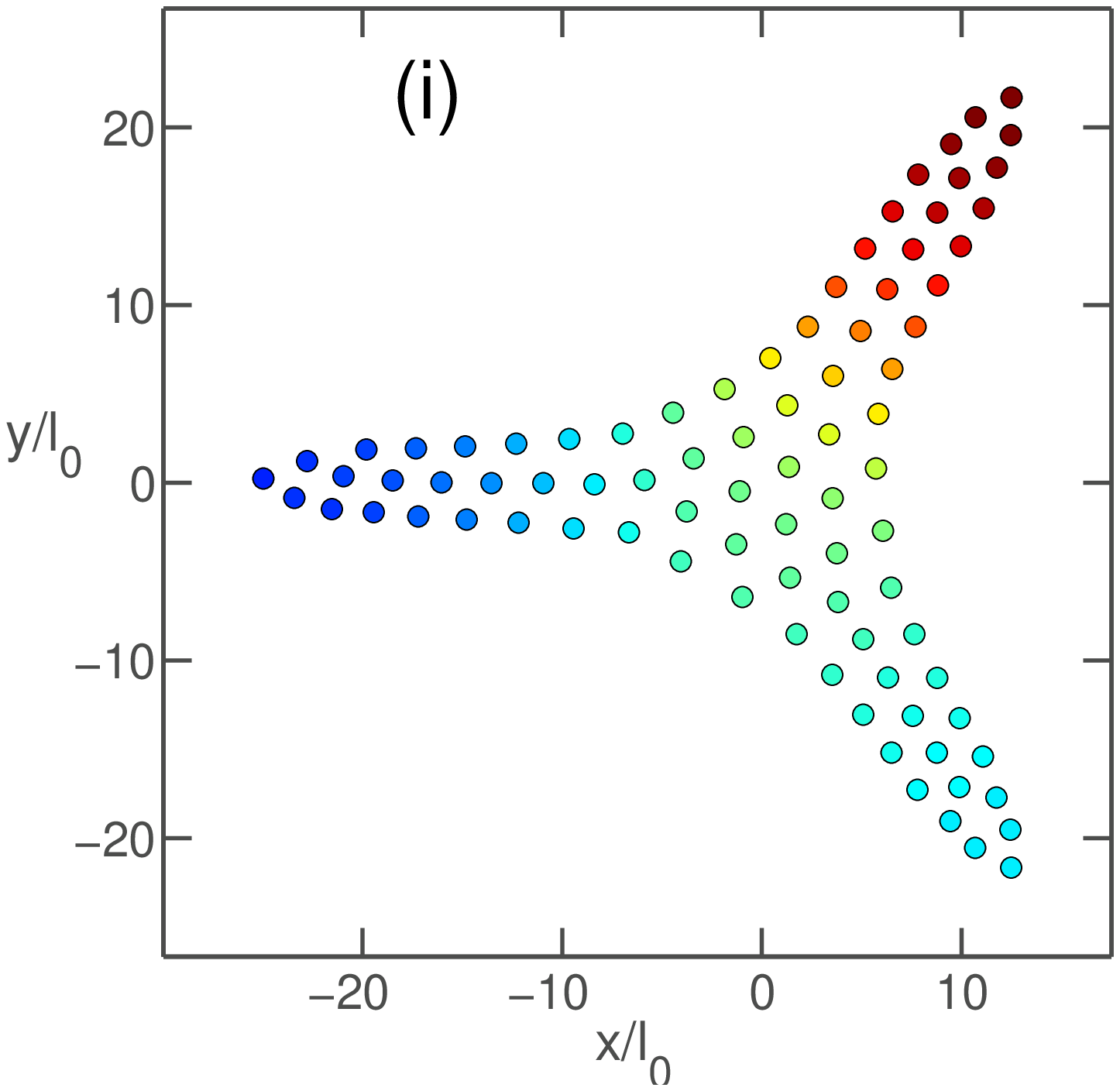}
\end{minipage}

\vspace{2mm}

\hspace{-11mm}
\begin{minipage}[b]{0.16\linewidth}
\includegraphics[scale=0.145]{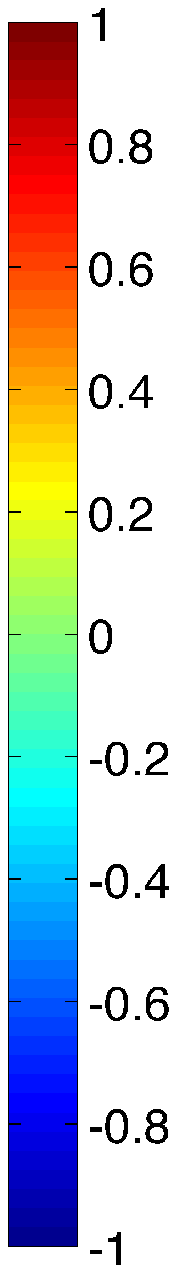}
\end{minipage}
\hspace{-5mm}
\begin{minipage}[b]{0.16\linewidth}
\includegraphics[scale=0.145]{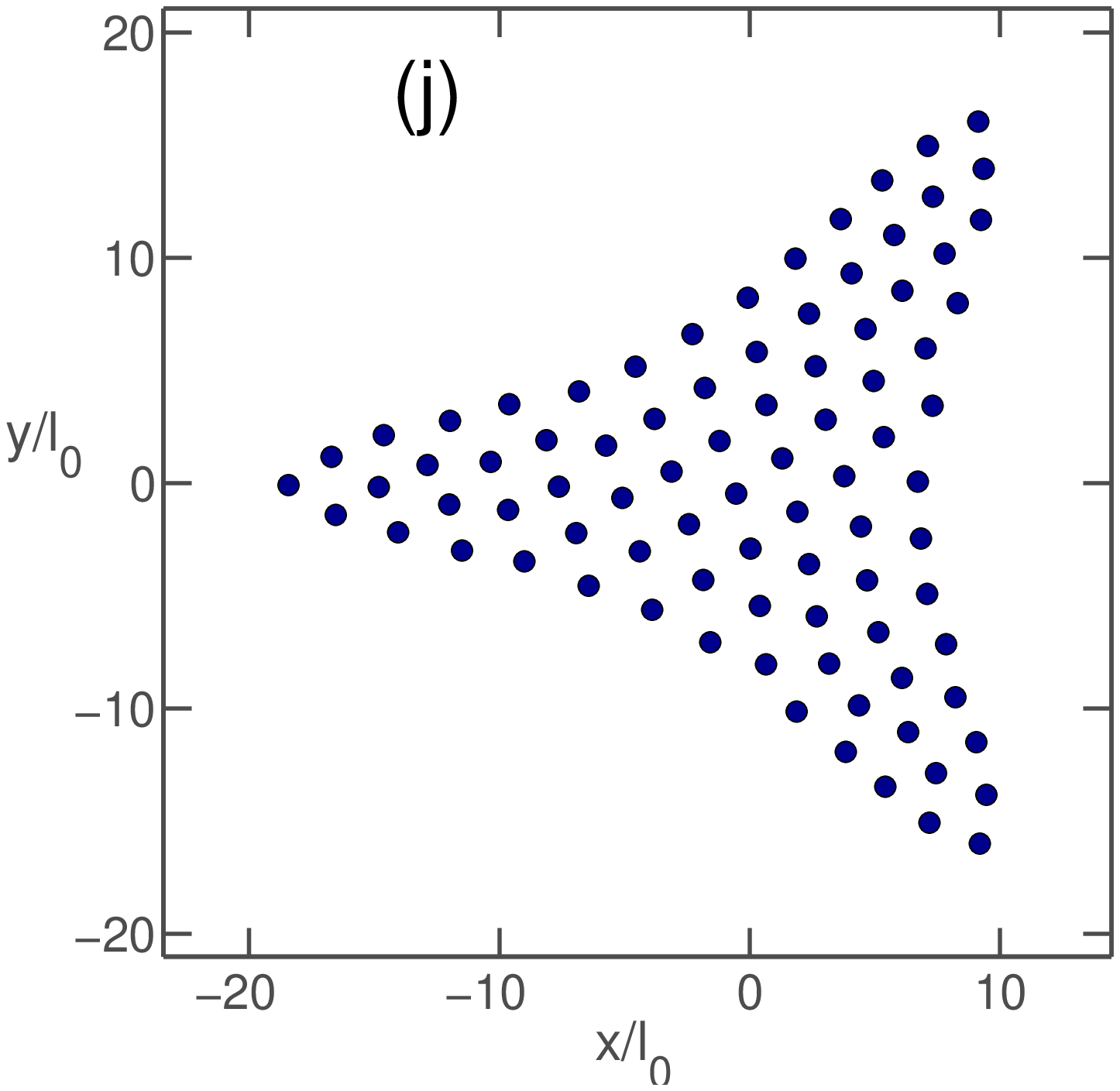}
\end{minipage}
\hspace{7mm}
\begin{minipage}[b]{0.16\linewidth}
\includegraphics[scale=0.145]{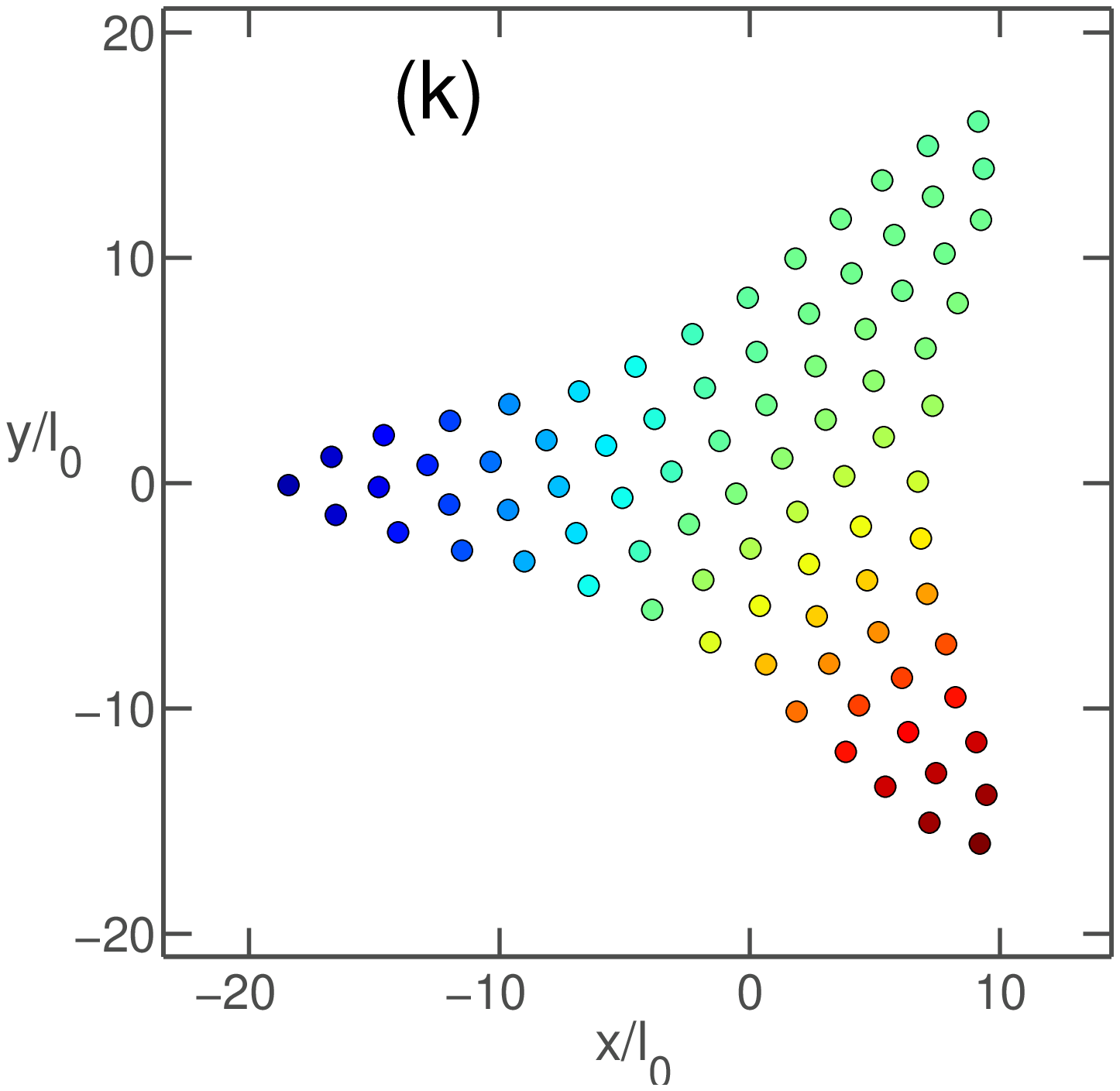}
\end{minipage}
\hspace{7mm}
\begin{minipage}[b]{0.16\linewidth}
\includegraphics[scale=0.145]{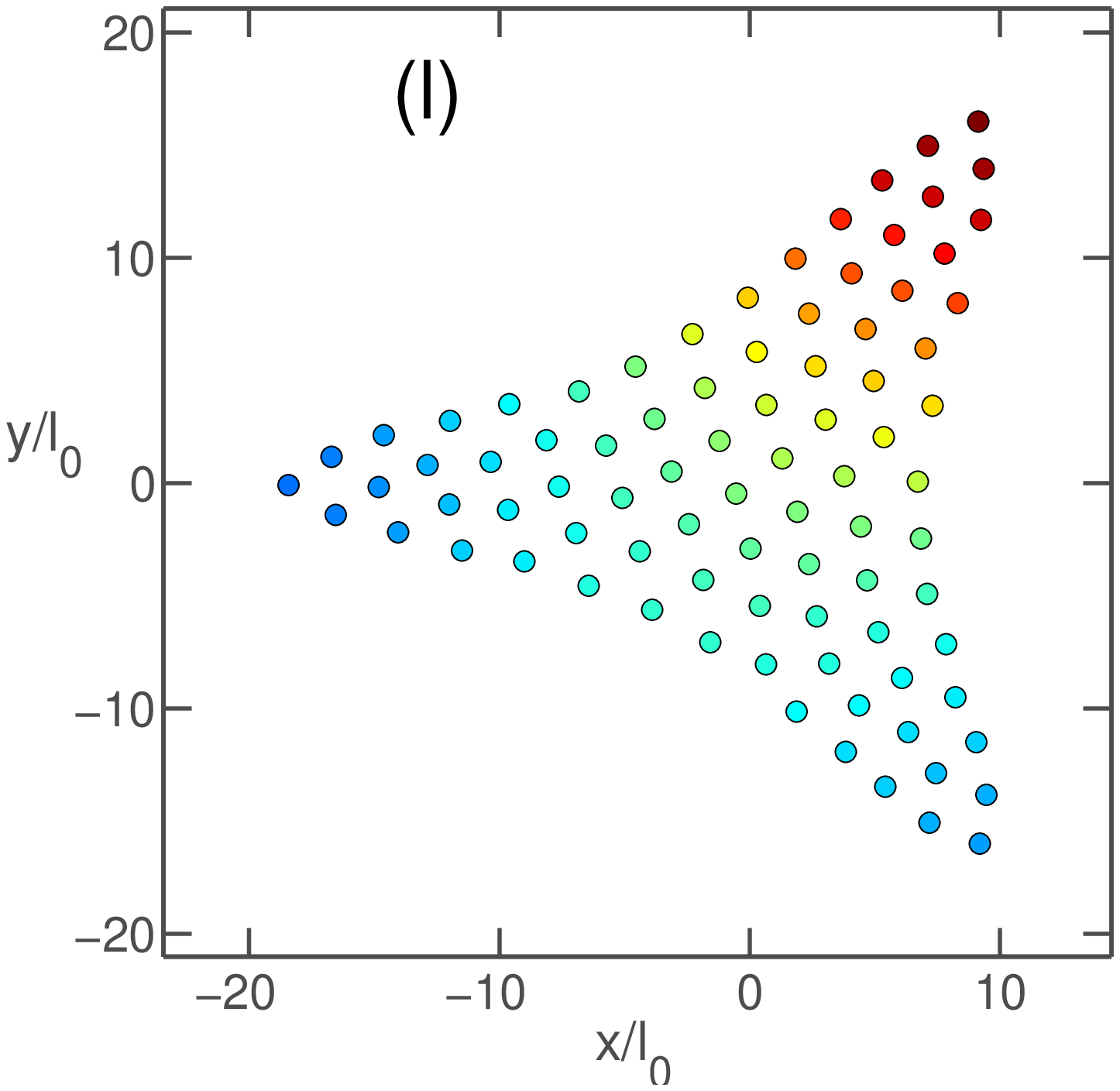}
\end{minipage}

\caption{(Color online.) Three highest-frequency axial eigenvectors for various trapping strengths $\omega_{eff}$ and rotating wall potentials $V_{W}$. Panels $(a)-(f)$ correspond to a low rotating wall strength $V_{W}=0.0025\omega_{z}^{2}$, $(a)-(c)$ representing eigenmodes of the crystal structure corresponding to $\omega_{eff}=0.21\omega_{z}$ and $(d)-(f)$ representing the crystal structure for $\omega_{eff}=0.24\omega_{z}$. Panels $(g)-(l)$ correspond to a stronger rotating wall strength $V_{W}=0.0040\omega_{z}^{2}$, $(g)-(i)$ corresponding to $\omega_{eff}=0.21\omega_{z}$ and $(j)-(l)$ corresponding to $\omega_{eff}=0.24\omega_{z}$.\label{fig: eigenvectors}}  

\end{figure}

Next, we discuss the eigenvectors of the axial mode phonons, which we can calculate immediately from the diagonalization of the stiffness matrix that yields the eigenfrequencies. Modes close to the center-of-mass mode are collective, where ions move with a long-wavelength. It is these long-wavelength modes that are important for purposes of quantum simulation, and we concentrate on these in our discussion here. In Fig.~\ref{fig: eigenvectors}, we show maximum axial displacements of the ions, corresponding to the highest three axial modes, with the displacements normalized to unity and color-coded as indicated in the adjacent colorbar. We do this for the two different strengths of the radial trapping strength $\omega_{eff}$, for both the high and the low rotating wall potentials $V_{W}$. The highest axial mode corresponds to the center-of-mass motion. 
The other two modes (so-called tilt modes) are nearly degenerate, and are seen to have similar nature even when anisotropy due to the rotating wall is dominant. This is also seen in Fig.~\ref{fig: normal_modes2}, where we see that the two modes just below the center-of-mass mode have the same eigenfrequencies. This behavior of the penultimate axial modes is different from the quadrupole-wall crystal, where this degeneracy of the modes is lifted under the corresponding anisotropic rotating wall, and an additional mode is sometimes introduced between them. 


Finally, we examine the strength of the effective spin-spin coupling that results from applying a spin-dependent dipole optical force detuned close to the axial phonon modes. We focus only on detuning frequencies $\delta = \mu- \omega_{z}$ to the blue of the center-of-mass mode ($\delta > 0$) where we expect to find a power law dependence of the spin exchange on the distance between the spins in the lattice~\cite{porras_cirac}. This behavior has been predicted and verified in experiments, and we expect it to hold even for this triangular-wall anharmonic crystal. In fact, the objective for making the lattice distances more uniform, is to also make the spin-spin couplings between adjacent spins more uniform, and this is would follow immediately if there exists a power law relation between the two ($J_{jj'}\simeq J_0/|\mathbf{R}_j^0-\mathbf{R}_{j'}^0|^\alpha$). 

\begin{figure}[ht]
\centering
\includegraphics[scale=0.45]{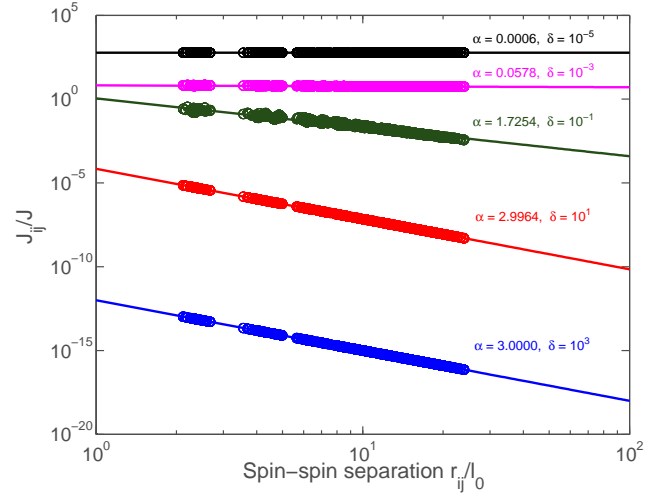}
\caption{(Color online.) Time-averaged spin-spin coupling coefficient $J_{ij}/J$ ($J=F_O^2/m\omega_z^2$) versus the distance between the ions $r_{ij}/l_0=|{\bf R}_i^0-{\bf R}_j^0|/l_0$ on a log-log plot. We plot the stable crystal structure exhibiting minimum variance in nearest-neighbor distances, corresponding to $V_{W} = 0.0025\omega_{z}^{2}$ and $\omega_{eff} = 0.25\omega_{z}$. The power law exponent $\alpha$ and the strength of the detuning away from the center-of-mass mode $\delta$ are both indicated near each curve. 
 \label{fig: J}}
\end{figure}

The parameters of axial phonon modes discussed already are used to calculate the static spin-spin interaction $J_{jj'}$ between the spins of ions $j$ and $j'$, based on Eq.~(\ref{eq: jij}). In Fig.~\ref{fig: J}, we plot this static interaction strength (expressed on a logarithmic scale) as a function of the distance between the ions, for various values of detuning $\mu$ larger than the center-of-mass frequency. We do this for the stable crystal structure exhibiting minimum variance in the nearest-neighbor distances. We see a behavior very similar to the quadrupole-wall potential. A uniform $J_{ij}$, independent of $r_{ij}$ indicates that the detuning laser excites only the uniform center-of-mass mode. As we move away from the center-of-mass mode, an increasingly large number of eigenmodes participate in the coupling, and we see a clear trend in the values, such that $J_{ij} \propto r_{ij}^{-\alpha}$. In the limit of large detuning, we have dipole-dipole interactions where $\alpha$ tends to a value of 3. For small detunings, we have the all-to-all case of $\alpha\rightarrow 0$. There are small departures from the power law behavior for intermediate detunings, while the very small and very large values of detuning show excellent agreement with the power law. Note that these results are indeed much more uniform than what was found for a a quadrupole rotating wall~\cite{britton}.

\begin{figure}[ht]
\centering
\includegraphics[scale=0.45]{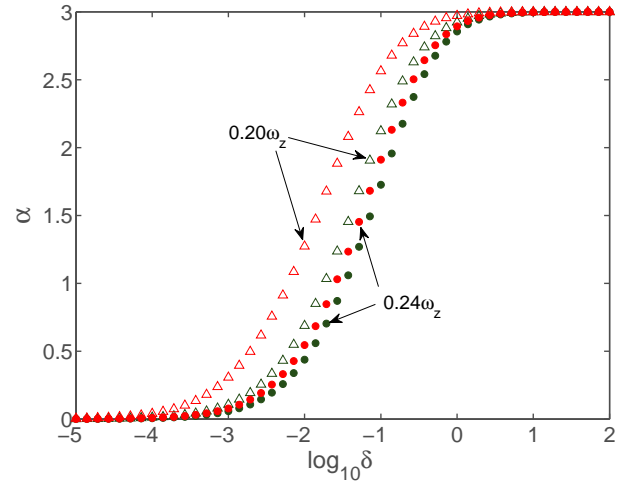}
\caption{(Color online.) Fitted exponent of the power law $\alpha$ (of the $J_{ij}$ coefficients as a functions of the distance $r_{ij}$) plotted against the strength of detuning away from the center-of-mass frequency, $\delta$, for the same set of trap parameters  and notation as in Fig.~\ref{fig: normal_modes2}. \label{fig: alpha}}
\end{figure}

In Fig.~\ref{fig: alpha}, we plot the fitted power-law exponent $\alpha$ versus the strength of the detuning away from the center-of-mass mode. The trend we see here is similar to the one seen in calculations for the quadrupole-wall potential. We see a faster approach to the dipole-dipole limit ($\alpha=3$) for both smaller effective (radial) trapping frequencies, and weaker strengths of the triangular rotating wall potential, just like for the quadrupole rotating wall. 

\begin{figure}[ht]
\centering
\includegraphics[scale=0.45]{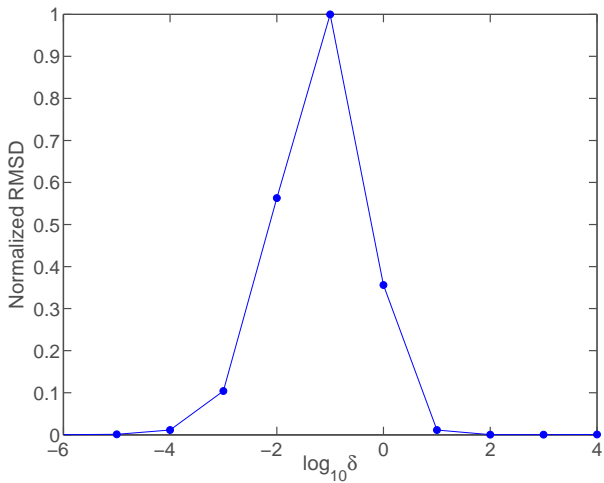}
\caption{Normalized oot mean square deviation for the fits of the spin coupling constants
to a power law for different detunings to the blue of $\omega_z$.  \label{fig: R}}
\end{figure}

We have already noted that there are deviations from the power law behavior for intermediate values of detuning, and this is apparent in the spread of values away from the linear fit in Fig.~\ref{fig: J} (especially for $\delta=10^{-1}\omega_{z}$). To explore these deviations in more detail, we plot the normalized root mean square deviation (RMSD), defined by,
\begin{equation}
{\rm normalized~RMSD}=\frac{\sqrt{\sum_{i<j}(J_{ij}-J_{ij}^{\rm fit})^2}}
{\max_\delta\sqrt{\sum_{i<j}(J_{ij}-J_{ij}^{\rm fit})^2}},
\label{eq: RMSD}
\end{equation}
as a function of the detuning $\delta$ in Fig.~\ref{fig: R}. We see an adherence to the power law (characterized by values of the  normalized RMSD close to 0) for both small and large detunings $\delta$. More importantly, we see the largest deviation in the normalized RMSD parameter for strengths of detuning in the intermediate range of $10^{-4}\omega_{z}$ to $10^{1}\omega_{z}$, We can understand this behavior easily if we look at the structure of the static part of Eq.~(\ref{eq: jij}), that relates the strength of the spin-spin coupling, $J_{ij}$, to the normal mode properties of the crystal. Each term in the summation can be understood to correspond to an eigenmode's contribution to the coupling strength. When the beat-note frequency $\mu$ is very close to $\omega_{z}$ ($\delta\approx 0$), only the center-of-mass mode (corresponding to uniform motion of the ions) contributes, and the $J_{ij}$ does not depend on distance. This behavior corresponds to a value of $\alpha=0$. For $\mu$ farther away from the center-of-mass mode, the lower modes begin to contribute increasingly. When only a few modes contribute, we cannot expect the power law behavior to hold~\cite{mcaneny_freericks}. The structure of the eigenvectors, as we see in Fig.~\ref{fig: eigenvectors}, is clearly not compatible with the power law decay of $J_{ij}$ with distance. For many ions sitting at opposite edges, there is a large $J_{ij}$ whereas the coupling is virtually zero for other pairs separated by much smaller distances. This is the origin of the spread of spin-spin couplings in Fig.~\ref{fig: J} and the increase in the value of the normalized RMSD in Fig.~\ref{fig: R}. As we move towards larger values of $\mu$ (and $\delta$), we see that all the modes begin to contribute almost equally, and in the limit of dipole-dipole interactions, we get a value of $\alpha$ very close to 3, with the normalized RMSD close to 0.

We note that such behavior is independent of the details of the crystal structure itself. The eigenvectors corresponding to the first few modes will have a structure independent of the details of the trap potential, and will cause a similar deviation from the power law behavior as we saw above. This deviation implies that the spin-spin couplings are no longer correlated with the distances between the ions, and hence an increase in spatial uniformity of the crystal is not guaranteed to have a bearing on the uniformity of the spin-spin interactions. This is an important observation, as the increasingly uniform nearest-neighbor distances for the triangular wall crystal would imply a more uniform spin-spin coupling strength only for detuning strength values that are moderately large. For intermediate values of $\delta$, it is important to consider the nature of modes just below the center-of-mass mode to describe the spin-spin coupling strength between ions corresponding to that strength of detuning.  
~

\section{\label{sec:level4}Conclusions and Discussion}

In this work, we have examined the properties of a Penning trap with an additional anharmonic and triangular rotating wall potential which provide a much more uniform ionic crystal for use in quantum simulation. By performing a detailed analysis of the equilibrium positions, the phonons, and the effective spin-spin interactions, we find that indeed one can generally obtain more uniform spin-spin coupling constants.  As one might have predicted, the relationship between ionic spacing in the lattice and the uniformity of the spin-spin interactions is not directly one-to-one.  For small $\alpha$ values, it is the character of the phonon eigenmodes that lie close to the center-of-mass mode that determine the behavior of the spin-spin couplings more than the interparticle spacing. 
We hope that the result of this work will be found to be useful in planning future experiments with the Penning trap platform that will employ additional anharmonic trap terms and a triangular rotating wall for a more uniformly spaced triangular lattice.

\acknowledgments

We acknowledge useful conversations with Dan Dubin and John Bollinger.
We acknowledge support from National Science Foundation under Grant No. PHY-1314295.
A. K. acknowledges support from the Indo-US Science and Technology Forum via the S. N. Bose Scholars Program.
J.K.F. further acknowledges support from the McDevitt
bequest at Georgetown University. B.Y. acknowledges
support from the Achievement Rewards for College Scientists
Foundation.

\end{document}